\documentclass[aps,superscriptaddress,amsfonts,amssymb,amsmath,showpacs,twocolumn,nofootinbib]{revtex4-1}
\usepackage[dvips]{graphicx}

\begin{document}
\title{Quasinormal modes for the scattering on a naked Reissner-Nordstr\"om singularity}

\author{Cecilia Chirenti} 
\email{e-mail: cecilia.chirenti@ufabc.edu.br} 
\address{Centro de Matem\'atica, Computa\c c\~ao e Cogni\c c\~ao, UFABC, 09210-170 Santo Andr\'e, SP, Brazil}
\author{Alberto Saa}
\email{e-mail: asaa@ime.unicamp.br}
\address{
Departamento de Matem\'atica Aplicada,
Universidade Estadual de Campinas,
13083-859 Campinas,  SP, Brazil}
\author{Jozef Sk\'akala}
\email{email: jozef.skakala@ufabc.edu.br}
\address{Centro de Matem\'atica, Computa\c c\~ao e Cogni\c c\~ao, UFABC, 09210-170 Santo Andr\'e, SP, Brazil}

\begin{abstract}
What should be the quasinormal modes associated with a
spacetime that contains a naked singularity instead of a black hole?
In the present work we address this problem by studying the scattering
of scalar fields on a curved background described by a
Reissner-Nordstr\"om spacetime with $|q| > m$. We show that there is a
qualitative difference between cases with $1 < q^2/m^2 \lesssim 9/8$ and
cases with $q^2/m^2 \gtrsim 9/8$. We discuss the necessary conditions for
the well-posedness of the problem, and present results for the low damped modes in the low
$l$ and large $l$ limit. We also consider 
 the asymptotically highly damped quasinormal modes. We present strong evidence that such modes are absent in the case of a naked Reissner-Nordstr\"om singularity, corroborating recent conjectures relating them to classical and quantum properties of horizons. 
\end{abstract}


\maketitle

\section{Introduction}

The naked Reissner-Nordstr\"om (R-N) singularity is a classical
general relativistic solution in electrovacuum. The solution is
expected to have a very limited meaning, due to the fact that such
singularities cannot be created neither by a gravitational collapse,
nor by dropping a charge into the black hole.
(According to the weak cosmic censorship conjecture general naked singularities
should be prohibited in general theory of relativity, although there
are indications that by including quantum effects the violations of
the conjecture could be considered \cite{Saa}.) Moreover, a naked
singularity created from some exotic initial data conditions should
become quickly neutralized (classically, or via quantum pair
production). Some results also indicate that if one considers
electro-gravitational perturbations the R-N naked singularity
becomes linearly unstable \cite{Dotti}.
However it was discovered that the scalar field scattering problem
on such a singular background can be still well defined \cite{Wald1,
Wald2, Sandberg, Gibbons, Horowitz, Pitelli}, since the waves remain
regular at the origin. (However, the back reaction of the given scalar field configuration might still excite some of the unstable electro-gravitational modes and this would eventually lead to a breakdown of the perturbation approach.) Despite the nice regularity property of the
scattering problem, the spacetime is non-globally hyperbolic and the
time evolution of the fields is not unique \cite{Ishibashi,
Martelini}. This means one has to specify an additional boundary
condition at the singularity to obtain a fully unique time
evolution. Another way of seeing the problem is through the language
of operators: one can understand the spatial part of the wave
operator as a positive symmetric operator acting on a $L^{2}$
Hilbert space, and then obtain the scalar field dynamics through a
suitable positive self-adjoint extension of such a symmetric
operator \cite{Wald1, Wald2}. (One ``preferred'' way in which such a
self-adjoint extension can be always realized is through the so
called Friedrich's extension \cite{Wald1}, which will also be the
case of this paper.) Anyway, after uniquely specifying the dynamics,
one should be able to characterize the scattering by a set of
characteristic oscillations, the quasi-normal modes.

Low damped quasi-normal modes are in general used as a possible
source of information about potential astrophysical objects (such as
neutron stars, black holes), and the highly damped modes are potentially
interesting from the point of view of quantum gravity \cite{Hod, Maggiore}.\footnote{For a paper dealing particularly with the R-N black hole case see \cite{Skakala}.} Since a lot
of work was devoted to the problem of quasinormal modes of the
Reissner-Nordstr\"om black hole, it might be interesting to observe
what happens if one transits from the R-N black hole case to the R-N
naked singularity case (with a reflective boundary condition).
Information about ``what happens'' shows how many features of the
quasinormal modes of the black hole spacetimes are specific to the
black holes themselves and what features survive much more general
conditions. Let us also give one concrete example why asking what happens with the quasinormal modes of the naked R-N singularity might be interesting: In the black hole case the behavior of the asymptotically highly damped modes is widely suspected to be linked to the properties of the black hole spacetime horizon(s), (or more specifically it is considered to carry information about \emph{quantum} black holes). What does then happen with the asymptotically highly damped modes in case there are \emph{no} horizons present? (If the behavior of the highly damped modes in case of R-N naked singularity would, for instance, resemble the behavior of the highly damped modes of the black hole, it would be a disturbing fact from the point of view of the popular conjectures linking the modes to the horizon's quantum area spacing. Moreover, investigating the R-N naked singularity from this point of view is especially attractive, as the R-N naked singularity is obtained by a continuous transition in the $q,~m$ parameters from the R-N black hole spacetime.)  Thus, briefly, we hope that despite the fact that
most likely the R-N naked singularity model does not correspond to a
realistic physical situation, there are still many interesting
things one can learn from such a model.  

The structure of this paper is as follows: In the second section we
analyse the problem of the uniqueness of the time evolution of
scalar fields on a R-N naked singularity background. In the sections
three and four we analyse the properties of the effective potential
for the scalar fields scattering and the geometrical optics
(eikonal) limit of such a scattering problem. In the fifth section
we define analytically solvable potentials that can give good
approximations to the problem of the low damped QNM frequencies
(such that characterize the given scattering problem). In the sixth
section we use those analytical approximations to derive
semi-analytical results for the QNM frequencies in the eikonal
limit. In the seventh section we use the numerical  characteristic
integration to obtain the low damped frequencies for the low values
of $l$. In the eighth section we analyse what happens with the asymptotically highly damped modes and we suggest that in case of naked R-N singularity such modes do \emph{not} exist (as one might expect considering some presently popular conjectures \cite{Hod,Maggiore}). We give the final conclusions in the section nine. We
provide also appendices with more detailed results and some further
suggestions for the analytical approximations of the problem.

\section{The time evolution problem for a scalar field in the R-N naked singularity}
\label{sec2}

In this section we will follow the standard analysis of the scalar field evolution in a curved background. (As an example of such an analysis see the treatment of Schwarzschild black hole perturbations in \cite{Schmidt, Leaver}. For a review that presents also such techniques see for example \cite{Nollert}.)  Take the Klein-Gordon equation for the complex (charged) scalar field:
\begin{equation}
\frac{1}{\sqrt{-g}}\partial_{\mu}(\sqrt{-g}~g^{\mu\nu}\partial_{\nu}\Psi)=0,
\end{equation}
with the metric line element given as
\begin{equation}
g_{\mu\nu}dx^{\mu}dx^{\nu}=-f(r)dt^{2}+f(r)^{-1}dr^{2}+r^{2}d\Omega^{2}.
\end{equation}
For the Reissner-Nordstr\"om (R-N) singularity the function $f(r)$ is in Planck units given as:
\begin{equation}\label{tortoise}
f(r)=1-\frac{2m}{r}+\frac{q^{2}}{r^{2}}, ~~~~(q^{2}>m^{2}).
\end{equation}
Take the decomposition of the field into the spherical harmonics
\begin{equation}
\Psi (t,r,\theta,\phi)=\sum_{l,m}\psi_{l}(t,r)Y_{ml}(\theta,\phi).
\end{equation}
After we separate the variables we obtain the following reduced equation
\begin{eqnarray}
\label{KGreduced}
\frac{d^{2}\psi_{l}(t,r)}{dt^{2}} &=&
 \frac{f(r)}{r^{2}}\frac{d}{dr}\left[ r^{2}f(r)\frac{d\psi_{l}(t,r)}{dr}\right]  \nonumber \\ 
 &-&\frac{l(l+1)f(r)}{r^{2}}\psi_{l}(t,r). 
\end{eqnarray}
We are interested only in compactly supported data initial value problem: first assume that $\psi_{l}(r,t)$ is everywhere bounded and hence the Laplace transform of $\psi_{l}(r,t)$ exists:
\begin{equation}
\tilde\psi_{l}(s,r)=\int_{0}^{\infty}dt~ e^{-st}\psi_{l}(t,r)
\end{equation}
The Laplace transformed equation \eqref{KGreduced} gives the following equation:
\begin{eqnarray}\label{KGreducedTransformed}
s^{2}\tilde\psi_{l}(s,r)&=&\frac{f(r)}{r^{2}}\frac{d}{dr}\left[ r^{2}f(r)\frac{d\tilde\psi_{l}(s,r)}{dr}\right] \nonumber \\
&-&\frac{l(l+1)f(r)}{r^{2}}\tilde\psi_{l}(s,r)+I_{l}(s,r),
\end{eqnarray}
where
\begin{equation}
I_{l}(s,r)=\left[s\psi_{l}(t,r)+\frac{d\psi_{l}(t,r)}{dt}\right]_{|t=0}.
\end{equation}
The solution that corresponds to the initial data term $I_{l}(s,r)$ is obtained by the inverse Laplace transform of $\tilde\psi_{l}(s,r)$ and the function $\tilde\psi_{l}(s,r)$ is given as
\begin{equation}
\tilde\psi_{l}(s,r)=\int_{0}^{\infty}dr' ~G_{l}(s,r,r')I_{l}(s,r').
\end{equation}
Here $G_{l}(s,r,r')$ is a Green's function satisfying
\begin{eqnarray}\label{Green's}
&-&\frac{f(r)}{r^{2}}\frac{d}{dr}\left[ r^{2}f(r)\frac{d G_{l}(s,r,r')}{dr}\right]  \\ &+&\left[s^{2}+\frac{l(l+1)f(r)}{r^{2}}\right] G_{l}(s,r,r')=\delta(r-r'). \nonumber
\end{eqnarray}
Since $\psi_{l}(r,t)$ was bounded, its Laplace transform $\tilde\psi_{l}(s,r)$ must be also bounded in $r$. This translates to the boundedness of the Green's function in $r$. The unique solution of the given initial data problem is obtained if the condition of boundedness of the Green's function leads to a \emph{unique} way to construct Green's function from the two linearly independent solutions $U_{l1}, U_{l2}$ of the homogeneous equation
\begin{eqnarray}\label{homogeneous}
&-&\frac{f(r)}{r^{2}}\frac{d}{dr}\left[ r^{2}f(r)\frac{d U_{l1,2}(s,r)}{dr}\right] \\ &+&\left[s^{2}+\frac{l(l+1)f(r)}{r^{2}}\right] U_{l1,2}(s,r)=0.\nonumber
\end{eqnarray}
If $f(r)$ goes to 1 at spatial infinity (the metric is asymptotically flat), then there is only one solution of \eqref{homogeneous} that stays bounded as $r\to\infty$. If there exists only one solution of the equation \eqref{homogeneous}, such that it is linearly independent from the solution bounded at infinity and in the same time it is bounded at $r\to 0$ then these two solutions uniquely define the Green's function. If all the solutions of \eqref{homogeneous} are singular at 0, there is no solution (bounded in $r$) of the given initial value problem such that it can be Laplace transformed (this can be taken as an indication that there is no solution at all).

If both of the linearly independent solutions are regular at 0, and (at least) two different Green functions lead to a function in the domain of the inverse Laplace transform, then there is no \emph{uniquely} defined solution to the initial value problem. In the "worst" case there are infinitely many solutions, given by arbitrary linear combination of $U_{l1}, U_{l2}$, that are linearly independent to the solution bounded at $r\to\infty$. In such a case the problem is underdetermined and one needs one more condition at $r=0$ that selects a unique Green's function between the different Green's functions marking different time evolutions.  For each one of the choices of the Green's function, one can reproduce the calculation from \cite{Schmidt, Leaver}
 and see that the quasinormal modes defined by
\begin{enumerate}
\item  the choice of the Green's function close to 0,
\item  the outgoing radiation condition,
\end{enumerate}
characterize the time evolution of the field at a fixed point within
some specific time interval. Unfortunately for the case of R-N naked
singularity ($f(r)=1-2m/r+q^{2}/r^{2}$,~~ $q^{2}>m^{2}$) both of the
linearly independent solutions $U_{l1}, U_{l2}$ are regular at 0 and
the problem is underdetermined. It is easy to show that both of the
solutions are regular at the origin. Write \eqref{homogeneous} as:
\begin{eqnarray}\label{homogeneous2}
&-&f(r)^{2}\frac{d^{2} U_{l1,2}(s,r)}{dr^{2}}-\left[\frac{f(r)(2r-2m)}{r^{2}}\right]\frac{d U_{l1,2}(s,r)}{dr}\nonumber\\ &+&
\left[s^{2}+\frac{l(l+1)f(r)}{r^{2}}\right] U_{l1,2}(s,r)=0.
\end{eqnarray}
Now taking the $r\to 0$ limit of the equation \eqref{homogeneous2}
one obtains the following
\begin{eqnarray}\label{homogeneous3}
-\frac{d^{2} U_{l1,2}(s,r)}{dr^{2}}&+&\left[\frac{2m}{q^{2}}\right]\frac{d U_{l1,2}(s,r)}{dr}\nonumber \\&+&\frac{l(l+1)}{q^{2}} U_{l1,2}(s,r)=0.
\end{eqnarray}
This means the solutions $U_{l1,2}$ behave close to 0 as
\begin{equation}\label{closezero}
U_{l1,2}(s,r)=\exp(\beta_{1,2}r),
\end{equation}
with
\begin{equation}
\beta_{1,2}=~\frac{m}{q^{2}}\pm\sqrt{\left(\frac{m}{q^{2}}\right)^{2}+\frac{l(l+1)}{q^{2}}}~~\in\mathbb{R},
\end{equation}
hence both are regular.

The fact that the problem is underdetermined  is not surprising,
since the space-time is \emph{not} globally hyperbolic and anything
can fall out at any time from the singularity. This means the
singularity has "hair" (carries some other information beyond the
metric) and the quasi-normal modes obviously depend on the ``hair".
Let us add here that, as we already mentioned in the introduction,
the ``hair'' of the singularity relates to the existence of many
different self-adjoint extensions of the ``Hamiltonian'' operator in
the equation. (In \cite{Ishibashi} one can find a nice analysis of
the uniqueness of the self-adjoint extensions of such operators for
many different types of naked spacetime singularities including the
R-N naked singularity.)

Is there any intuitive physical condition that we can further impose on the fields, that will uniquely select the appropriate Green's function? At least to get the geometrical optics continuous extension of the black hole case one can impose the condition that nothing falls in or out of the singularity. This means there is neither absorption nor superradiation in the scattering and the S-matrix of the K-G field is a unitary operator. What does this condition mean?
The conserved current 4-vector for the complex Klein-Gordon field is given by
\begin{equation}\label{current}
J^{\mu}(\Psi)=-i g^{\mu\nu}\left(\Psi\bigtriangledown_{\nu}\Psi^{*}-\Psi^{*}\bigtriangledown_{\nu}\Psi\right).
\end{equation}
Let us integrate the 4-current along a cyllindrical hypersurface given by $r=r_{0}$ and $t\in[t_{1}, t_{2}]$. Then since \eqref{current} is a conserved current it holds that
\begin{equation}\label{currentcondition}
Q(t_{2})-Q(t_{1})=-\int_{\Sigma_{r_{0}}}~dt ~d\theta ~d\phi~ \sqrt{-h}~n^{\mu}J_{\mu}.
\end{equation}

Here $Q(t_{1,2})$ is the integral along a hypersurface given by the interior of the cylinder at the constant time ($t_{1}$ or $t_{2}$), with a future orientated surface normal vector. $\Sigma_{r_{0}}$ is a cylindrical hypersurface given by $r=r_{0}$ and  $\sqrt{-h}$ is an induced density given as $\sqrt{-h}=\sqrt{f(r_{0})}~r_{0}^{2}\sin(\theta)$. Furthermore $n^{\mu}$ is a normal vector to $\Sigma_{r_{0}}$ given as $n^{(t,r,\theta,\phi)}=(0,\sqrt{f(r_{0}}),0,0)$. Since we want $Q$ to remain constant with respect to time as we take the limit $r_{0}\to 0$ (nothing flows out or into the singularity), the equation \eqref{currentcondition} reduces to the following
\begin{widetext}
\begin{equation}\label{integral}
\lim_{r_{0}\to 0}\int_{\Sigma_{r_{0}}}dt ~d\theta ~d\phi~ \sqrt{-h}~n^{\mu}J_{\mu}= -i\lim_{r_{0}\to 0}f(r_{0})r_{0}^{2}\int^{t_{2}}_{t_{1}}dt\int_{0}^{\pi} d\theta \sin\theta \int_{0}^{2\pi} d\phi   [\Psi\partial_{r}\Psi^{*}-\Psi^{*}\partial_{r}\Psi] =0.
\end{equation}
\end{widetext}
Let us analyse the equation \eqref{integral} in  decomposition into spherical harmonics, hence
\begin{equation}
\Psi(t,r,\theta,\phi)=\sum_{l,m}\psi_{l}(r,t) Y_{lm}(\theta,\phi).
\end{equation}
 Then we can rewrite the equation \eqref{integral} as
\begin{eqnarray}\label{finalcondition}
&\ & \lim_{r_{0}\to 0} f(r_{0})r_{0}^{2}\sum_{l}\int_{t_{1}}^{t_{2}} dt \left[\psi_{l}(r,t)\partial_{r}\psi_{l}^{*}(r,t) \nonumber \right.  \\
&-& \left.  \psi^{*}_{l}(r,t)\partial_{r}\psi(r,t)\right]_{|r=r_{0}} =0,
\end{eqnarray}
for arbitrary $t_{1}, t_{2}$. Since $f(r_{0})r^{2}_{0}\to q^{2}$ as
$r_{0}\to 0$, in order to fulfill the equation
\eqref{finalcondition} we impose\footnote{Such a condition on the
radial part of the current at the origin (determining whether the
scattering is absorptive, radiative, or superradiative) occurs in
the formulation of \cite{Martelini}, where it gives constrains on
the domain of the ``Hamiltonian'' providing symmetricity of the
operator.} for every $l$ and every $t$
\begin{equation}\label{fixingcond}
J_{r}(\Psi,0)=\left[\psi_{l}(r,t)\partial_{r}\psi_{l}^{*}(r,t)-\psi^{*}_{l}(r,t)\partial_{r}\psi_{l}(r,t)\right]_{|r=0}=0.
\end{equation}
But this means that the functions $\psi_{l}(r,t)$ should be always constrained either by the condition $\psi_{l}(0,t)=0$ or $\partial_{r}\psi(r,t)|_{r=0}=0$. (One might argue that it will be enough to claim that the fields and their first $r$ derivatives should be real at zero, but that does not put any general constraint on the normal modes.)
Now we obtain the function $\psi_{l}(r,t)$ from the normal modes as
\begin{equation}
\psi_{l}(r,t)=\int_{0}^{\infty}d\omega ~e^{-i\omega t}\left[c_{1}(\omega)U_{l1}(r,i\omega)+c_{2}(\omega)U_{l2}(r,i\omega)\right],
\end{equation}
and take the condition $\psi_{l}(0,t)=0$ for arbitrary time $t$. This translates to
 \begin{equation}\label{Foulike}
\int_{0}^{\infty}d\omega ~e^{-i\omega t}\left[c_{1}(\omega)U_{l1}(0,i\omega)+c_{2}(\omega)U_{l2}(0,i\omega)\right]=0.
\end{equation}
But since both $U_{l1}(0,i\omega)=U_{l2}(0,i\omega)=1$ (see  \eqref{closezero}) and Fourier-like transform given by \eqref{Foulike} should not map non-zero functions to zero, one obtains the condition $c_{2}(\omega)=-c_{1}(\omega)$.
This means we compose the relevant wave packet only from the following modes:
\begin{equation}\label{modes1}
\tilde U_{l}(r,i\omega)=U_{l1}(r,i\omega)-U_{l2}(r,i\omega).
\end{equation}
The same line of reasoning applies to the condition $\partial_{r}\psi(r,t)|_{r=0}=0$ and the wave-packets that fulfill such condition must be composed entirely from modes given as ($l>0$):
\begin{equation}\label{cond2}
\tilde
U_{l}(i\omega,r)=\frac{U_{l1}(i\omega,r)}{\beta_{1}}-\frac{U_{l2}(i\omega,r)}{\beta_{2}}.
\end{equation}
For $l=0$, we have the coefficient $\beta_{2}=0$ and the solution
$U_{l2}(r,i\omega)$ behaves as a constant for $r\approx 0$ (with
the value set to 1), which means that the condition \eqref{cond2}
does not make strictly sense; in such case the modes are given
simply by the function $U_{l2}(r,i\omega)$.

This shows that the condition of S-matrix being a unitary operator
gives additional constraints on the Green's function. The previous
conditions mean that we shall consider only a linear space of
wave-packets formed purely from modes that vanish at zero, or a
linear space of wave-packets formed purely from modes whose first
$r$ derivatives vanish at zero. (Of course one cannot superpose
wave-packets formed from modes having the vanishing radial
derivative at 0 with wave packets formed from modes that are
vanishing at 0, since the 4-current does not linearly depend on a
wave-function.)

So let us pick one of those two types of modes, fix the particular
Green's function and employ the following reasoning: Since $J_{0}$
is conserved and the coefficients in the equation \eqref{KGreduced}
are everywhere outside 0 a smooth function of $r$, the function
$\psi(r,t)$, arising from compactly supported initial data could be
unbounded in time only in the case the wave-packet becomes
concentrated around the singularity and slowly growing
asymptotically with time into delta function
\begin{equation}
\lim_{t\to\infty}\psi(r,t) \sim\delta(r).
\end{equation}
But this scenario is prevented by the boundary condition that ``nothing flows in or out of the singularity at any time''. This means we expect the solution to be bounded with respect to both time and space and all the reasoning based on the assumption of the existence of Laplace transform is justified.

Further in the text we will employ the field vanishing condition at
0. (This boundary condition at the singularity corresponds to what
is known as Friedrich's extension of a symmetric operator.) Thus the
quasinormal modes will relate to the scattering problem following
from the time evolution determined by the boundary condition
$\psi(0,t)=0$. But everything that we will do in the following text
can be repeated for any other meaningful normal modes boundary
condition giving another time evolution of the fields. The general
dependence of the quasi-normal modes (characteristic oscillations)
on such a boundary condition (given by some linear combination
$U_{l1}(r,i\omega)+K U_{l2}(r,i\omega)$) describes the way the
quasi-normal modes depend on the ``hair" of the singularity. Since
the quasi-normal modes are the ones that carry astrophysical
information about the astrophysical sources, their dependence on the
``hair" of the singularity (in some simplifying sense given by the
complex parameter $K$) has a potential astrophysical importance (of
course in case naked singularities have any astrophysical
importance).

\section{The scalar wave scattering on a naked singularity}

Using $\phi_{l}$ defined as $\phi_{l}(r,t)=r\psi_{l}(r,t) $ and $x$ the tortoise coordinate given by the condition:
\begin{equation}
\frac{dr}{dx}=f(r),
\label{x}
\end{equation}
one can rewrite the equation \eqref{KGreduced} into the following form
\begin{equation}\label{wave}
\frac{\partial^{2}\phi_{l} (x,t)}{\partial
t^{2}} - \frac{\partial^{2}\phi_{l} (x,t)}{\partial
x^{2}} = V(m,q,l,x)\phi_{l}(x,t),
\end{equation}
with
\begin{equation}\label{potential}
V(m,q,l,x)=\left[\frac{l(l+1)}{r^{2}(x)}+\frac{2m}{r^{3}(x)}-\frac{2q^{2}}{r^{4}(x)}\right]
f\big(r(x)\big).
\end{equation}
And, for the normal modes ~$e^{-i\omega t}\phi_{l}(r)$, we can write
\begin{equation}\label{R-W}
\frac{\partial^{2}\phi_{l} (x)}{\partial
x^{2}}+\left[\omega^{2}-V(m,q,l,x)\right]\phi_{l}(x)=0.
\end{equation}

\begin{figure}[!htb]
  \includegraphics[angle=270,width=\linewidth]{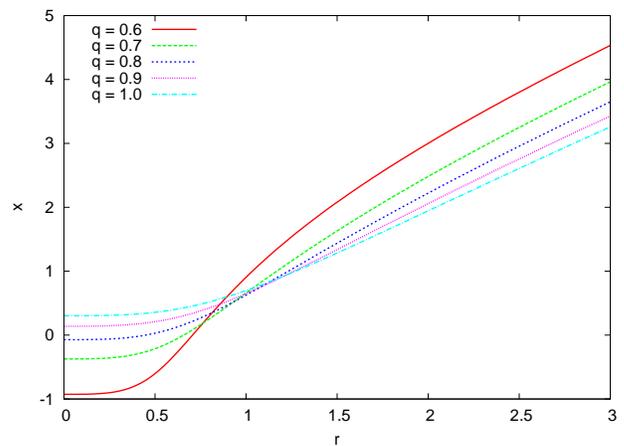}
\caption{Typical examples of the behavior of the new tortoise coordinate defined by eq.~(\ref{rstarnew}) for a spacetime with $m = 0.5$ and different values of $q$, taking $C = 0$.}
\label{fig:rstar}
\end{figure}

If $|q|>m$ , we can see that $f(r)$ given by eq.~(\ref{tortoise}) has
no zeros for real arguments, but eq.~(\ref{x}) can still be
integrated to give
\begin{eqnarray}
x = r &+& \frac{2m^2 - q^2}{\sqrt{q^2-m^2}}\arctan\left( \frac{r-m}{\sqrt{q^2-m^2}}\right) 
\nonumber \\
&+& m\ln(r^2 - 2mr + q^2) + C\,,
\label{rstarnew}
\end{eqnarray}
where $C$ is an integration constant. We remark here $r_* \to
\infty$ as $r \to \infty$, but for $r\to 0$ we have
\begin{eqnarray}
x(r \to 0)  &=& \frac{2m^2 - q^2}{\sqrt{q^2-m^2}}\arctan\left( \frac{-m}{\sqrt{q^2-m^2}}\right)\nonumber \\ &+& m\ln(q^2) + C= x(0) = \textrm{constant},
\end{eqnarray}
as can be seen in fig.~(\ref{fig:rstar}). Further in the text we
take the tortoise coordinate with the boundary condition $x(0)=0$.
It means that with respect to the usual ($C=0$) tortoise coordinate
such tortoise coordinate is shifted to the origin by the
transformation $x\to x(r)-x(0)$.

\begin{figure}[!htb]
\begin{center}
  \includegraphics[angle=270,width=1\linewidth]{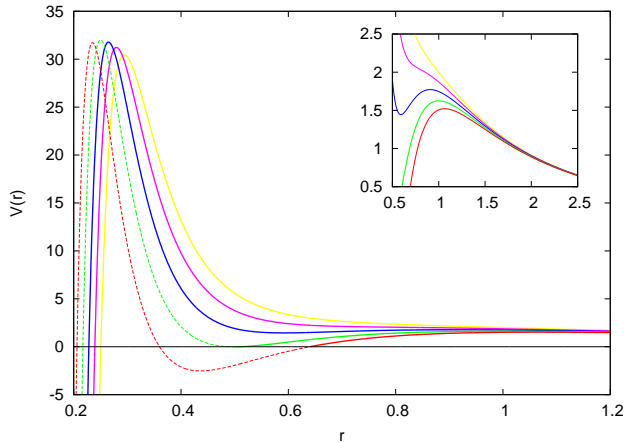}
\end{center}
\caption{Potential $V(r)$ given by eq.~(\ref{potential}) with $l=2$, $m = 0.5$ for $q = 0.48, 0.5, 0.52, 0.54$  and 0.56. (The curves from left to right correspond to the increase of charge.) Note that the dashed part of the potential (for $q = 0.48$ and 0.5) is inside the black hole horizon.}
\label{fig:pot}
\end{figure}

The potential \eqref{potential} has for the ratio $q^{2}/m^{2}$ less than
approximately $9/8$ and the relevant $x$ (in the naked singularity
case the domain of $x$ is constrained) 3 extrema, one smaller ``outer''
maximum, one dominant ``inner'' maximum and minimum in the potential
valley between them. (For $r\to 0$ the function $V(r)\to -\infty$.) For $q^{2}/m^{2}$ more than approximately $9/8$ the potential has only one maximum (thus only one peak). These features of the potential \eqref{potential} can be seen in the figure \ref{fig:pot}.

\begin{figure}[t]
\begin{center}
  \includegraphics[angle=270,width=\linewidth]{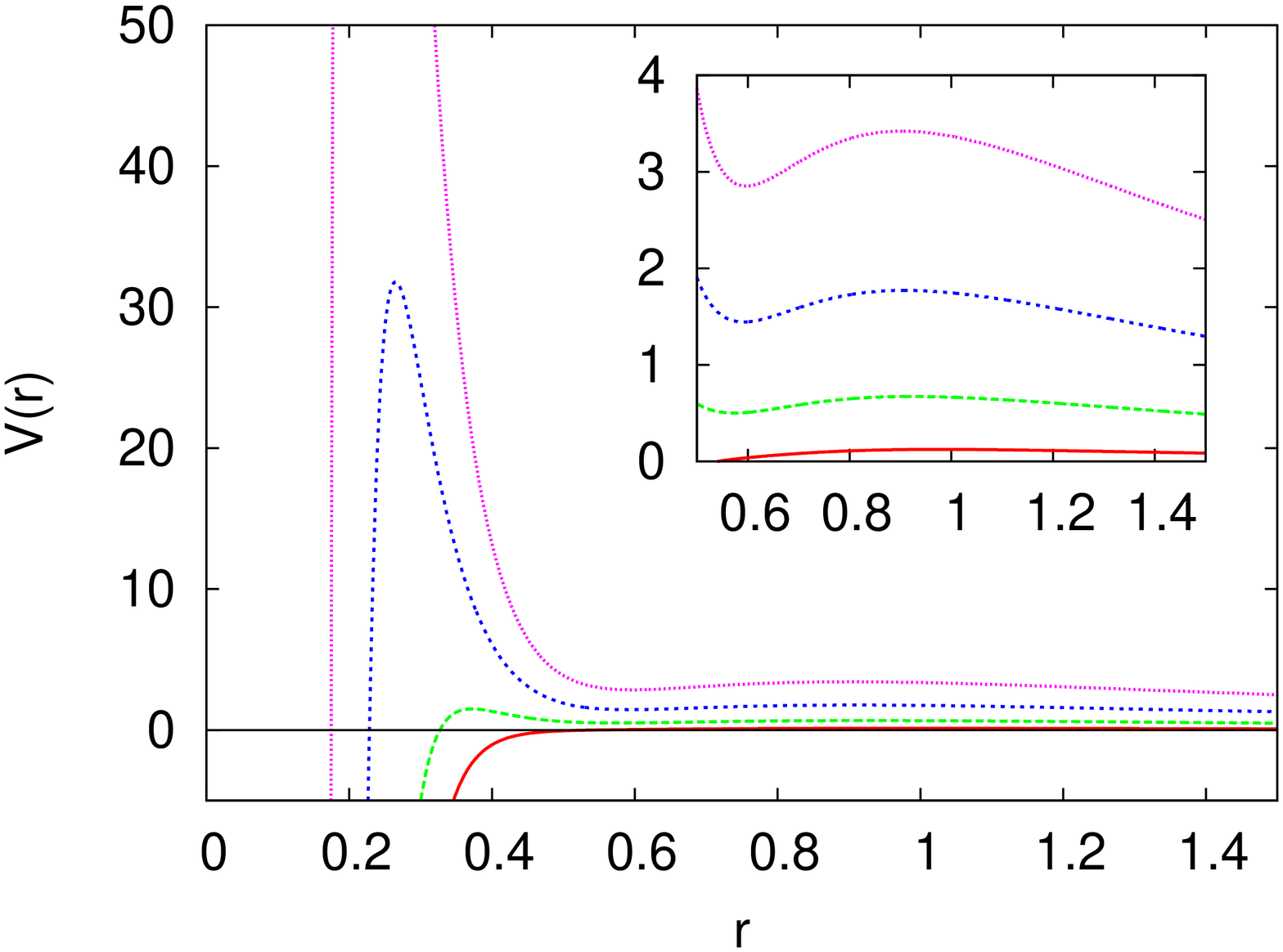}
  \includegraphics[angle=270,width=\linewidth]{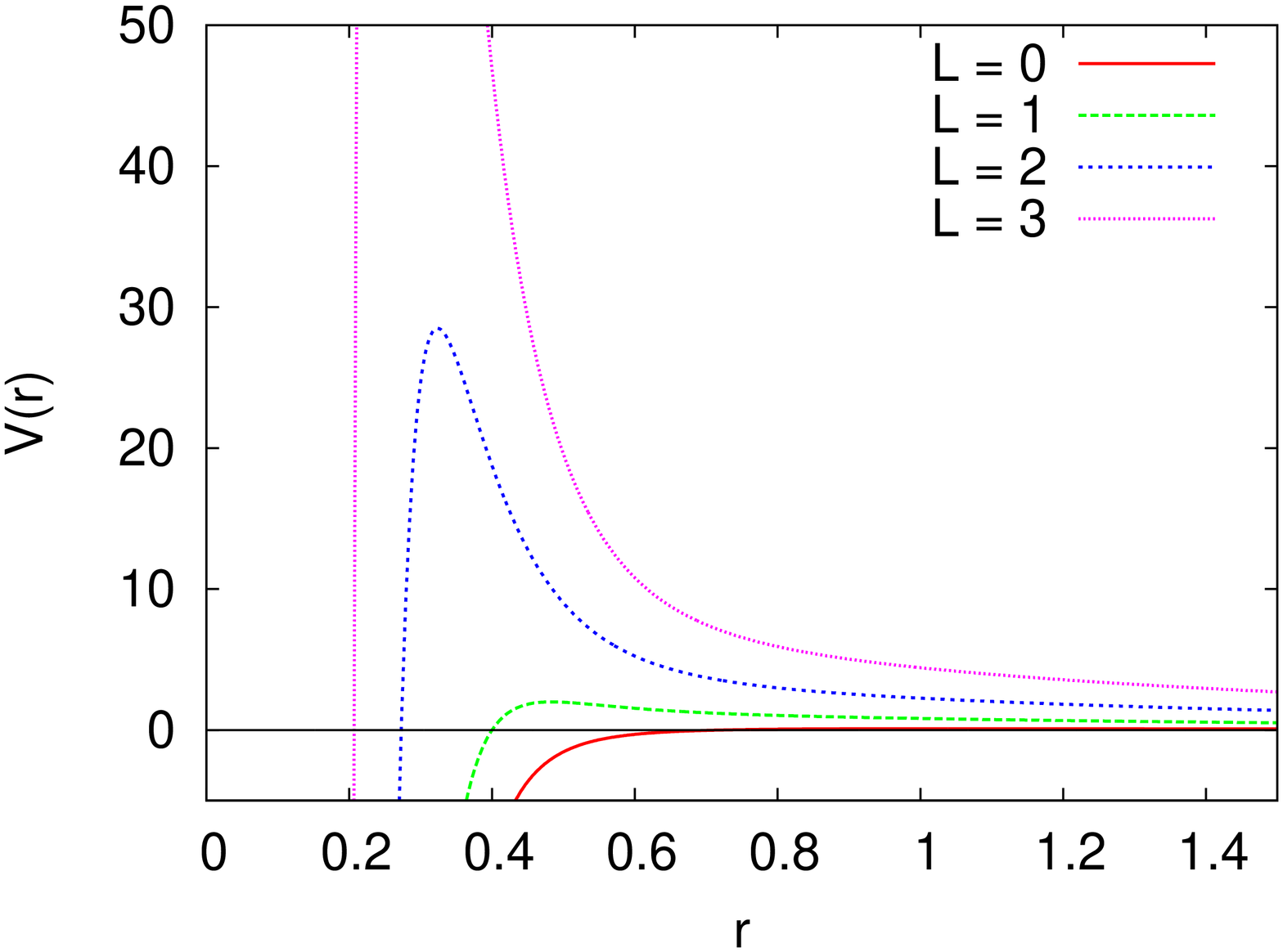}
\end{center}
\caption{Above: potential $V(r)$ given by eq.~(\ref{potential}) with $m = 0.5$ for $q = 0.52$ ($q^2/m^2 < 9/8$) and different values of $l$ (same caption as on the right plot). Note in this case the existence of the secondary peak, reminiscent of the black hole potential. Below: the same, but this time for $q = 0.6$ ($q^2/m^2 > 9/8$). There is no secondary peak in this case.}
\label{fig:pot_L}
\end{figure}

Moreover, in the case  $q^{2}/m^{2}$ less than
approximately $9/8$ and $l\gg 1$, the inner maximum becomes completely dominant,
 making the
outer peak negligible as compared to the size of the inner peak. For
the outer peak and for $l\gg 1$ the terms
$\frac{2m}{r^{3}}-\frac{2q^{2}}{r^{4}}$
represent only a small correction (as compared to the term
proportional to $l^{2}$) and the second peak will vanish in this
approximation at\footnote{As trivially expected, we will see that the same result comes directly from the potential for the motion of a massless particle.} $q^{2}/m^{2}=9/8$.
(This is because the outer peak lies always at $r>1$. It is also
quite obvious that the inner peak exists only due to the fact that
the terms $\frac{2m}{r^{3}}-\frac{2q^{2}}{r^{4}}$ become dominant
for $r$ close to 0.)  The features described in this paragraph can be observed in the figure \ref{fig:pot_L}.

Further in the text we will use the following notation related to the potential parameters: By $V_{1}, V_{2}$ we mean the heights of the two peaks.
($V_{1}$ is the first larger peak, $V_{2}$ the second smaller peak.)
By $\alpha_{1}$ we call the curvature of the first peak and by
$\alpha_{2}$ the curvature of the second peak. Furthermore, by $x_{1max}$ we
mean the point of the location of the top of the first peak and by
$x_{2max}$ the point of the location of the second peak. Many of these parameters can be exactly calculated in the $l\gg 1$ limit. The results of these calculations are given in the appendix \ref{B}.

\section{Eikonal limit - the particle picture}

First let us have a look at the geometrical optics (eikonal) limit of the
Klein-Gordon equation, which is valid for $l\gg 1$.  In particular
what does the geometrical optics limit tell us about the fundamental
mode? The effective potential for the massless particle is
\begin{equation}
V(r)=\frac{l^{2}}{r^{2}}f(r).
\end{equation}
The effective potential goes to plus infinity as $r\to 0$ and its
extremes are given as
\begin{equation}
r_{1,2}=\frac{3m\pm m\sqrt{9-8\frac{q^{2}}{m^{2}}}}{2}.
\end{equation}
This means the potential has one minimum and one maximum for
\begin{equation}\frac{q^{2}}{m^{2}}<\frac{9}{8}.\end{equation}
For
\begin{equation}\frac{q^{2}}{m^{2}}\geq\frac{9}{8}\end{equation}
there are no extrema. The maximum represents the unstable orbit of a
massless particle, and for the black hole case it is only the
maximum that is relevant (because it is located above the horizon). For the
naked singularity with $q^{2}/m^{2}\leq 9/8$ there exists also a
stable orbit of a  massless particle. However the fundamental mode
in the geometrical optics limit is related to the unstable orbit as
\begin{equation}\label{geometrical}
\omega=\sqrt{V(r_{max})}-\frac{i}{2}\sqrt{-\frac{V''(r_{max})f^{2}(r_{max})}{2
V(r_{max})}}.
\end{equation}
(For a very good paper that discusses this topic see
\cite{Cardoso}.) Note that here
$\sqrt{-\frac{V''(r_{max})f^{2}(r_{max})}{2 V(r_{max})}}$ gives the
unstable orbit decay rate and furthermore holds the following
\begin{eqnarray}
V''(r_{max})f^{2}(r_{max})&\equiv & \left(\frac{\partial^{2}
V(r)}{\partial r^{2}}f^{2}(r)\right)|_{r_{max}}\nonumber \\ &=&\frac{\partial^{2}
V(x)}{\partial x^{2}}|_{x_{max}}.
\end{eqnarray}
(Again $x$ is the tortoise coordinate.)
Particularly for the Reissner-Nordstr\"om black hole/naked
singularity (with  $q^{2}/m^{2}< 9/8$ ) equation \eqref{geometrical}
can be expressed as:
\begin{widetext}
\begin{equation}
\omega=\frac{l}{r_{max}}\sqrt{1-\frac{2m}{r_{max}}+\frac{q^{2}}{r^{2}_{max}}}-
\frac{i}{2r_{max}}\sqrt{-3+\frac{18m}{r_{max}}-\frac{13q^{2}+24m^{2}}{r^{2}_{max}}+\frac{32mq^{2}}{r^{3}_{max}}-\frac{10q^{4}}{r^{4}_{max}}}
\end{equation}
\end{widetext}
with $r_{max}=\frac{3m+\sqrt{9m^{2}-8q^{2}}}{2}$, given as before.

From the black hole QNM-s boundary conditions one can see that such
massless particles decaying from their unstable orbit and eventually
escaping either to the B-H horizon, or the infinity represent the
fundamental QNM-s in the $l\gg 1$ (geometrical) limit. One can observe
(see the discussion later) that the same holds in the naked
singularity case. In the $l\gg 1$ case the fundamental mode of the
massless perturbations is described by the picture of a particle
decaying from its unstable ``photon'' orbit. (Note that in the naked
singularity case the stable photon orbit  makes the particle
decaying inwards oscillate around the stable orbit.) This should
suggest that in the geometrical optics limit
\begin{enumerate}
\item one shall expect continuity of the fundamental mode (as a function of $q, m$) when turning from the black hole to the naked singularity case,
\item for $q^{2}/m^{2}\geq 9/8$ there is a clear indication that there do not exist low damped quasi-normal modes.
\end{enumerate}

The way how to understand the continuity (for $l\gg 1$) of the fundamental mode as one transfers from the black hole case to the naked singularity case could be the following: Consider the partial S-matrix $S_{l}(\omega)$. Within the S-matrix
one can look for poles on the complex plane: obviously any of the
poles must be necessarily non-real. For the relativistic time
dependence convention chosen as ~$\exp (-i\omega t)$~ the poles with
$\omega_{I}>0$ must be purely imaginary (as a result of the given
Hamiltonian being a symmetric operator) and they represent bound
states with the non-relativistic energies given as
$E=-\omega^{2}_{I}$. The poles with $\omega_{I}<0$ represent
quasinormal frequencies. Some of the quasinormal frequencies
correspond to the resonances in the phase factor related to the partial S-matrix, and the
wave packets formed out of the resonance energies represent quantum
particles tunnelling out of the potential valley; for the
non-relativistic quantum particles (for example) with the decay rate
given as $(\omega^{2})_{I}=2\omega_{R}\omega_{I}$. Now
 in the naked singularity
case the fundamental mode is represented by a resonance that has a low enough energy
(at the level of the smaller peak) and gets trapped in the potential
valley in between the peaks for some time and then radiated away.
The larger peak behind the smaller peak is hugely dominant (large
$l$) and effectively acts to the wave packet as an infinite
barrier. There are probably much higher resonance energies related
to higher QNM frequency overtones that are determined by the higher
peak, but left unaffected by the details of the small peak. So there
are two different effective regimes for the quasinormal modes, in
the first regime they are sensitive only to the details of the smaller
peak, in the second regime they are sensitive only to the details of
the larger peak. This splitting disappears in the case of small
$l$-s since the two peaks are of comparable heights and effectively
interfere.

Let us finish this section with one more remark: In the black hole
case one can relate the fundamental mode to the peak of the
potential (independently of $l\gg 1$) following the way of thinking in
\cite{Will}. The ratio of the amplitudes of the reflected and
transmitted waves $A_{ref}(\omega)/A_{trans}(\omega)$ is for
quasinormal modes 1. The same happens for real $\omega$-s, when the
energies $\omega^{2}$ are close (``almost at'') to the peak of the
potential. So for ``almost'' real $\omega^{2}$ (fundamental mode)
the continuity of the ratio of reflected and transmitted amplitudes 
analytically continued to the complex plane suggests that the real
part of the frequency should be near the square root of the peak of
the potential. (This intuition is then to some extend confirmed also
semi-analytically, such as by the Ferrari and Mashhoon approach
using the Poeschl-Teller potential \cite{Ferrari}, or by using the
inverted harmonic oscillator potential.) In the case of a naked
singularity life is not so easy (in the reduced 1D problem we have
no transmitted amplitudes), so this simple logic fails. (One can say
luckily it fails as if it remained valid one would also for $l\gg 1$
expect a discontinuity in the fundamental mode as one jumps from the
black hole to the naked singularity. This is because the second,
dominant peak suddenly appears in the domain of the $x$ coordinate,
due to the fact that the domain of $x$ discontinuously jumps when
passing from the black hole to the naked singularity case.)
Certainly one can say that any of the quick fits, such as were done
in the black hole case do not lead to anything close to the
numerical data obtained in this paper.

\section{General analytical methods to model the problem}

In this section we want to suggest some ways to model the naked singularity scattering (with a general time evolution) by solutions of the equation \eqref{R-W} with $V(x,l,m,q)$ replaced by analytically solvable potentials. The reasons are the following: First, this section serves as a basis for the analysis of the ~$l\gg 1$~ cases provided in the section \ref{largelsection}. Second, it demonstrates that the naked singularity scattering is \emph{in principle} treatable via analytical approximations and thus shows the general power of analytical techniques. (Also the approximations obtained here might produce some future results for the quasinormal frequencies. For some further calculations see the appendix \ref{A}. Furthermore, for a very nice overview of the results for the quasi-normal frequencies and related transmission resonances of the analytically tractable potentials see \cite{Boonserm}.)
Third, it might bring more insights into the physics obtained through the exact scalar field potential \eqref{potential}.

The key point is to split the scalar field potential into different
domains and approximate it on each of those domains (or directly the
solution of the equation \eqref{R-W}) by a different, analytically
solvable potential. Then one has to impose the standard procedure:
The logarithmic derivatives of the solutions on different domains
must be glued on the domain's boundary. We might add here that since
the analytically solvable potentials are typically fitted by the
parameters of the peaks one expects such approximation to work for
the modes that are \emph{not} too damped (the \emph{low} damped QNM
modes). The approximation \emph{cannot} be taken too seriously for
the \emph{highly} damped modes.

Let us suggest modeling all the solutions of the relevant cases as
follows:

\subsection{The case $q^2/m^2 \gtrsim 9/8$  for arbitrary $l$ \label{one}} For this case it might be
interesting to consider the Morse potential. (For the definition and
origins of all the potentials used in this section see also
\cite{Boonserm}.) On the other hand the infinite valley behind the
one and only peak may not be well modelled by only a finite valley
that is behind the peak of the Morse potential. Thus we consider the
solution in the region `behind'' Morse potential region to be given
as \eqref{modes1} for $x\to 0$ and the logarithmic derivative gluing
condition originates from:
\begin{enumerate}
\item   $0<x<a_{1}$~: use function given as (for the details see
equation \eqref{modes1}):
\begin{equation}\label{modes2}
\phi(r)=r\Psi(r)=A r(e^{\beta_{1}r}-e^{\beta_{2}r}).
\end{equation}
\item    $a_{1}\leq x$~:  use the appropriate solution of the equation \eqref{R-W} with the Morse
potential:
\[V_{1}e^{-\alpha_{1}(x-x_{1max})}\left(2-e^{-\alpha_{1}(x-x_{1max})}\right).\]
\end{enumerate}
Here $a_{1}$ is a point most conveniently chosen where the
Morse potential is 0 (for any arbitrary $l$ there is such point for
$x>0$). This means $a_{1}$ can be analytically given as
\begin{equation}
a_{1}=x_{1max}-\frac{1}{\alpha_{1}}\ln(2).
\end{equation}
Let us make here one remark: Instead of approximating the original scalar field equation for
$r\approx 0$ one can consider taking the equation \eqref{R-W} in the
$r\to 0$ approximation. This leads to the following:
\begin{equation}\label{phi}
\frac{d^{2}\phi(x)}{dx^{2}}+\frac{2}{9}\frac{\phi(x)}{x^2}=0, ~~~~x\to
0.
\end{equation}
The equation \eqref{phi} has solutions $\phi=C_{1} x^{1/3}+C_{2}
x^{2/3}$. The solution such that fulfills $\Psi(0)=0$ must have $C_{1}=0$. This
approximation of the solution we think to be less exact as the
approximation given as \eqref{modes2}, as it is in fact effectively only the first non-zero term of
the power series expansion of \eqref{modes2} taken at 0. (For such reasons we decided to use in our analytical approximations the approximate solution \eqref{modes2}.)

\subsection{The case of $q^2/m^2 \gtrsim 9/8$ in the large $l$ limit\label{two}} 

Let us consider now  the limit of
large $l$. In this case, since the peak already lies close to 0
and the change in the highest power of inverted $x$ becomes more and
more dominant (with higher $l$-s), one might simplify the
approximation given in case \ref{one}, by using the following
potential:
\begin{enumerate}
\item   $0<x<x_{1max}$:  use function given as \eqref{modes2}
\item    $x_{1max}\leq x$:   use the appropriate solution of the equation \eqref{R-W} with the
potential (Poeschl-Teller potential):
\begin{equation}V_{1}\cosh^{-2}[\alpha_{1}(x-x_{1max})].\end{equation}
\end{enumerate}
Here we also used the Poeschl-Teller potential instead of the Morse
potential, because the logarithmic derivative gluing condition is
much easier to solve for the solutions of the Poeschl-Teller than
for the solutions of the Morse potential. The expense of the simplicity is that the approximation by such
potential might be slightly less exact than by using the Morse potential, but one still expects it to be accurate enough.

\subsection{The case $q^2/m^2 \lesssim 9/8$
for arbitrary $l$} 
\label{three}
If one does not want to neglect
the second smaller peak (so one is interested in resonances related
to the valley between the two peaks), the analytically treatable
potential describing all the resonances (and also higher damped QNM
frequencies) could be:
\begin{enumerate}
\item   $0<x<a_{1}$:  use function given as \eqref{modes2},
\item    $a_{1}\leq x\leq a_{2}$~:   solutions of the equation \eqref{R-W} with the Morse potential 
\begin{equation}V_{1}e^{-\alpha_{1}(x-x_{1max})}\left(2-e^{-\alpha_{1}(x-x_{1max})}\right)\end{equation}
\item    $a_{2}\leq x$:   solution of the equation \eqref{R-W} with the Poeschl-Teller
potential:
\begin{equation}V_{2}\cosh^{-2}[\alpha_{2}(x-x_{2max})].\end{equation}
\end{enumerate}
Here $a_{1}$ is best chosen as in the case \ref{one} and
$a_{2}$ is chosen to be such that the resulting potential is
continuous. Generally $a_{2}$ has to be obtained numerically, but
for larger $l$-s (certainly $l=20$ is more than enough, as we checked), the curvature of the second peak is very small comparing
to the curvature of the first peak and also the curvature of the
first peak becomes (for large $l$-s) very large comparing to the
scales of the potential, so that one can calculate (for $l=20$ with
a good approximation at least to the 6 decimal places) $a_{2}$ just
by
\begin{enumerate}
\item  taking the Poeschl-Teller potential to be constant and given by the hight of the outer peak
\item simplifying the Morse potential by the following approximation:
\begin{equation}V(x)\approx 2V_{1}e^{-\alpha_{1}(x-x_{1max})}.\end{equation}
\end{enumerate}
 The resulting formula then becomes
\begin{equation}
a_{2}=x_{1max}-\frac{1}{\alpha_{1}}\ln\left(\frac{V_{2}}{2V_{1}}\right).
\end{equation}
Then one can easily show that $a_{2}\to 0$ as $l\to\infty$.

\subsection{The case of $q^2/m^2 \lesssim 9/8$
in the large $l$ limit\label{four}} Now consider what happens with
the resonances related to the valley between the two peaks in the
large $l$ limit. In such case the dominant peak grows to infinity as
compared to the smaller peak, the infinite valley behind the
dominant peak shrinks to 0. Also $a_{1}\to 0$ and the valley between
the two peaks becomes flat as compared to the difference between the
height of the first peak and the value of the potential at the
bottom of the valley. If the resonances related to the smaller peak
locate close to the top of the peak (they should since the first
peak grows with $l$ an makes the tunnelling harder), then one might
effectively approximate the case \ref{three} by the Poeschl-Teller
potential:
\begin{equation}
V_{2}\cosh^{-2}[\alpha_{2}(x-x_{2max})],       
\end{equation}
for $0\leq x$.
(One can also derive this approximation straight from the logarithmic derivative gluing condition in the appendix \ref{section} by taking the $l\to\infty$ limit.)

\section{The large $l$ limit - the analytic approach\label{largelsection}}

In this section we want to analytically confirm the results obtained directly from the eikonal limit.

\subsection{Calculations for $q^{2}/m^{2}<9/8$}

The case \ref{four} from the previous section is easily solvable. The solution in the
region $x\geq 0$ must fulfil the outgoing radiation condition. (To
be exact after analytically extending the solution to the complex
plane it must give asymptotically the outgoing waves on the line
$(\omega x)_{I}=0$.) Such solution is given as follows
\begin{widetext}
\begin{equation}\label{solution1}
\phi_{R}(x)=C_{2}e^{i\omega x}   F_{21}\left(g_{1},g_{2},g_{3},
(1+\exp(2\alpha_{2}(x-x_{2max})))^{-1}\right), 
\end{equation}
where $F_{21}$ is the standard hypergeometric function (of the type
2-1) and
\begin{eqnarray}
g_{1}&\equiv&\frac{1}{2}+\sqrt{\frac{1}{4}-\frac{V_{2}}{\alpha_{2}^{2}}}\equiv -g_{2}+1
\\
g_{3}&\equiv& 1-\frac{i\omega}{\alpha_{2}}.
\end{eqnarray}
The boundary condition at 0 gives $\phi_{R}(0)=0$, leading to
\begin{equation}\label{condition1}
F_{21}\left(\frac{1}{2}+\frac{i\sqrt{V_{2}}}{\alpha_{2}},\frac{1}{2}-\frac{i\sqrt{V_{2}}}{\alpha_{2}},1-\frac{i\omega}{\alpha_{2}},(1+\exp(-2\alpha_{2}x_{2max}))^{-1}\right)=0
\end{equation}
\end{widetext}
Unless $q^{2}/m^{2}$ is not too close to the upper limit given for
$l$ large as 9/8, one can reliably approximate
$e^{-2\alpha_{2}x_{2max}}\approx 0$ and the condition
\eqref{condition1} turns to be:
\begin{eqnarray}\label{condition2}
F_{21}\left\{\frac{1}{2}+\frac{i\sqrt{V_{2}}}{\alpha_{2}},\frac{1}{2}-\frac{i\sqrt{V_{2}}}{\alpha_{2}},1-\frac{i\omega}{\alpha_{2}},1\right\}=0
\end{eqnarray}

Now \eqref{condition2} can be rewritten through Gamma functions and
reduces to the simple problem of finding poles of the  
 product of Gamma functions 
\begin{equation}
\Gamma\left[-\frac{i\omega-i\sqrt{V_{2}}}{\alpha_{2}}+\frac{1}{2}\right]\Gamma\left[-\frac{i\omega+i\sqrt{V_{2}}}{\alpha_{2}}+\frac{1}{2}\right]=0.
\end{equation}
The poles are located at $-n$ for $n$ being a natural number and
this gives
\begin{equation}\label{formula}
\omega=\pm\sqrt{V_{2}}-i\alpha_{2}\left(n+\frac{1}{2}\right).
\end{equation}
This is precisely the formula for the lowest damped black hole QNM
frequencies. This means in the $l\gg 1$ limit we see (at least for the
lowest modes) a continuous transition from the black hole case to
the naked singularity case, as expected. (At least in the case where it
holds that $e^{-2\alpha_{2}x_{2max}}\approx 0$, which becomes a less
accurate approximation for $q^{2}/m^{2}$ close to $9/8$. On the
other hand for such ratios of $q^{2}/m^{2}$ the peak is almost
vanished and one can assume that the decaying circular orbit eikonal picture has already
quite limited sense.)

All this means that in the regime $l\gg 1$ the lowest QNM frequencies
should be located close to the first peak of the potential and are
(up to certain $n$) given by the formula \eqref{formula}. They
represent resonances such that they are insensitive to the details
of the larger peak as in their case the larger peak can be already
seen to act effectively as an infinite potential barrier. Hence we
can conclude that in the $l\gg 1$ case we are able to match the
geometrical (eikonal) limit of the original problem considered.

\subsection{Calculations for $q^{2}/m^{2}\geq 9/8$}

Take the case \ref{two} from the previous section. On the left hand side of the logarithmic derivative
gluing condition we need a logarithmic derivative of the function:
\begin{equation}
r\Psi_{L}=C_{L}r\left(e^{\beta_{1}r}-e^{\beta_{2}r}\right).
\end{equation}
On the right hand side of the logarithmic derivative gluing condition we
need the logarithmic derivative of the solution on the interval
$[x_{1max},\infty]$. It is given as in \eqref{solution1}, only the
parameters are related to the first and only peak in this case. Then
the logarithmic derivative gluing condition leads to the following:
\begin{widetext}
\begin{equation}
\label{first}
\frac{\Gamma\left(-\frac{i\omega}{2\alpha_{1}}-\frac{1}{2}\sqrt{\frac{1}{4}-\frac{V_{1}}{\alpha_{1}^{2}}}+\frac{3}{4}\right)\Gamma\left(-\frac{i\omega}{2\alpha_{1}}+\frac{1}{2}\sqrt{\frac{1}{4}-\frac{V_{1}}{\alpha_{1}^{2}}}+\frac{3}{4}\right)}{\Gamma\left(-\frac{i\omega}{2\alpha_{1}}-\frac{1}{2}\sqrt{\frac{1}{4}-\frac{V_{1}}{\alpha_{1}^{2}}}+\frac{1}{4}\right)\Gamma\left(-\frac{i\omega}{2\alpha_{1}}+\frac{1}{2}\sqrt{\frac{1}{4}-\frac{V_{1}}{\alpha_{1}^{2}}}+\frac{1}{4}\right)}= 
\frac{f(r_{max})}{2\alpha_{1}}\left[\frac{1}{r_{max}}+\frac{\beta_{1}e^{\beta_{1}r_{max}}-\beta_{2}e^{\beta_{2}r_{max}}}{e^{\beta_{1}r_{max}}-e^{\beta_{2}r_{max}}}\right].
\end{equation}
If one takes the $l\to\infty$ limit then the equation \eqref{first}
becomes:
\begin{equation}\label{tilde}
\frac{\Gamma\left(-i\tilde\omega+\frac{1}{2\sqrt{6}}+\frac{3}{4}\right)\Gamma\left(-i\tilde\omega-\frac{1}{2\sqrt{6}}+\frac{3}{4}\right)}{\Gamma\left(-i\tilde\omega+\frac{1}{2\sqrt{6}}+\frac{1}{4}\right)\Gamma\left(-i\tilde\omega-\frac{1}{2\sqrt{6}}+\frac{1}{4}\right)}=\frac{1}{4}\left[\frac{1}{\sqrt{3}}+\coth(\sqrt{3})\right].
\end{equation}
\end{widetext}
Here $\tilde\omega$ is defined as
$\tilde\omega=\omega/(2\alpha_{1})$. For not highly damped
frequencies in the $l\gg 1$ limit it must hold that
$\tilde\omega\in\mathbb{R}$. But by plotting the right side of the equation \eqref{tilde} minus the left side of the equation \eqref{tilde} in Mathematica it can be shown that the equation
\eqref{tilde} does not have any real solutions.
 For more than large intervals of values
of $\tilde\omega$ the absolute value of the left hand side of the
equation minus the right hand side of the equation always seems to grow (almost)
linearly with respect to $\tilde\omega$ from the value approximately given as 0.35 at $\tilde\omega=0$ to infinity. It might
not be hard to prove also analytically that \eqref{tilde} does not
have real solutions by using many relatively simple properties of
Gamma functions.
 This means there are \emph{no} low damped modes in
the case of $q^{2}/m^{2}>9/8$ and $l\gg 1$, as expected from the
eikonal limit / particle picture. (It means the damping of the
fundamental mode grows to infinity as $l\to\infty$ and grows no less
rapidly than $\alpha_{1}$, hence no less rapidly than cubically with $l$.)

\section{The naked singularity for the small wave mode numbers - numerical results for the frequencies}

In \cite{Alberto}, the quasi-normal modes of a scalar field in an electrically charged Vaidya background were studied using a similar numerical setup to what we are going to use here. An interesting investigation done in \cite{Alberto} tried to determine what would happen with the quasi-normal modes as the time-dependent background approached a naked singularity, but the numerical code used was not suitable for following the field evolution after the extremal configuration ($q=m$) was reached. In our present work, we have a static configuration that describes a naked a singularity in order to study the properties of the quasi-normal modes.

\begin{figure}[!htb]
\begin{center}
  \includegraphics[angle=270,width=\linewidth]{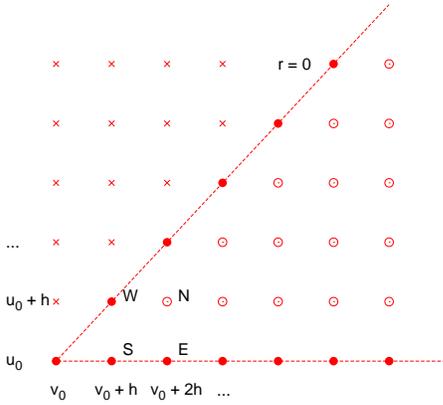}
\end{center}
\caption{Example of the numerical grid used for integrating eq.~(\ref{waveuv}) with boundary conditions (\ref{BC1}) and (\ref{BC2}). The points marked with ``x'' are out of our domain, values of $\phi$ at the positions marked with filled circles are given by the boundary conditions and the values at the empty circles are obtained with the algorithm (\ref{algo}).}
\label{fig:grid}
\end{figure}

Our objective in this section is to solve eq.~(\ref{wave}) with potential~(\ref{potential}) numerically, in the case where $q>m$ as described in the last section. To do this, we rewrite eq.~(\ref{wave}) in terms of the light-cone variables $u = t - x$ and $v = t + x$, where $x$ corresponds to the tortoise coordinate (\ref{x}), as
\begin{equation}
\frac{\partial^2 \phi}{\partial t^2} - \frac{\partial^2 \phi}{\partial x^2} = -
4\frac{\partial^2\phi}{\partial u \partial v} = V(r)\phi\,,
\label{waveuv}
\end{equation}
that can be integrated with the boundary conditions
\begin{eqnarray}
\label{BC1}
\phi(r=0,t)&=&\phi(u,v=u+2x_{0}) = 0\,,\\
\phi(u=0,v)&=&e^{-\frac{(v-v_c)^2}{2\sigma^2}}\,,
\label{BC2}
\end{eqnarray}
where condition~(\ref{BC1}) is a necessary condition on the field $\phi$ near the origin (see the discussion on fig. \ref{fig:phi_r} below) and condition~(\ref{BC2}) defines an ``arbitrary'' relevant initial signal to be propagated. We use the algorithm
\begin{equation}
\phi_N = \phi_W + \phi_E - \phi_S - \frac{\phi_W+\phi_E}{8}V\Delta_v\Delta_u\,,
\label{algo}
\end{equation}
see where $\Delta_u$ and $\Delta_v$ are the integration steps in $u$ and $v$, respectively, and the definitions of $\phi_N$ etc can be seen on fig.~ \ref{fig:grid}. Note that here $V$ is the potential (\ref{potential}) evaluated at the same $r$ coordinate as $\phi_S$ (and $\phi_N$).

As we can see in fig. \ref{fig:phi_r}, the boundary conditions (\ref{BC1}) and (\ref{BC2}) ensure the necessary conditions on the fields $\phi$ and $\psi$ near the center. As we discussed previously in section \ref{sec2}, the physically correct boundary condition for $\psi$ is $\psi(0,t) = 0$. From this we must have for $\phi(r,t) = r\psi(r,t) $ that $\phi(0,t) = 0$ and $\phi'(0,t) = 0$.
In fact, the quantity
\begin{equation}
{\cal E} = \int \left[\left(\frac{\partial \phi}{\partial t}\right)^2 +
\left(\frac{\partial \phi}{\partial x}\right)^2 +V(r)\phi^2
\right] dx
\end{equation}
is invariant along the $t$-evolution governed by (\ref{waveuv}) with the boundary conditions
$\phi(r=0,t)= \phi(r\to\infty,t)=0$. Taking into account (\ref{x}), the integrand can be expressed near $r=0$ as
\begin{equation}
\frac{q^4}{r^4}\left(\left(\phi'\right)^2 - \frac{2}{r^2}\phi^2\right),
\end{equation}
with the prime denoting the derivative with respect to $r$. Any initial condition with finite
${\cal E}$, as those ones obeying (\ref{BC1}) and (\ref{BC2}), is such that
\begin{equation}
\left(\phi'\right)^2 - \frac{2}{r^2}\phi^2 \sim r^{2+\alpha},
\end{equation}
for $r\to 0$,
with $\alpha>0$, implying that $\phi'(r=0,t)=0$ for initial condition with finite {\cal E} obeying $\phi(r=0,t)=0$. Moreover,
we have checked that our boundary conditions for $\phi$ also reproduce the $\phi'(0,t) = 0$ condition, by testing the code with the condition $\phi(\bar{r},t) = 0$, where $\bar{r}$ is a very small value, and we found no differences in the late time evolution of the fields.

\begin{figure}[!htb]
\begin{center}
  \includegraphics[angle=270,width=\linewidth]{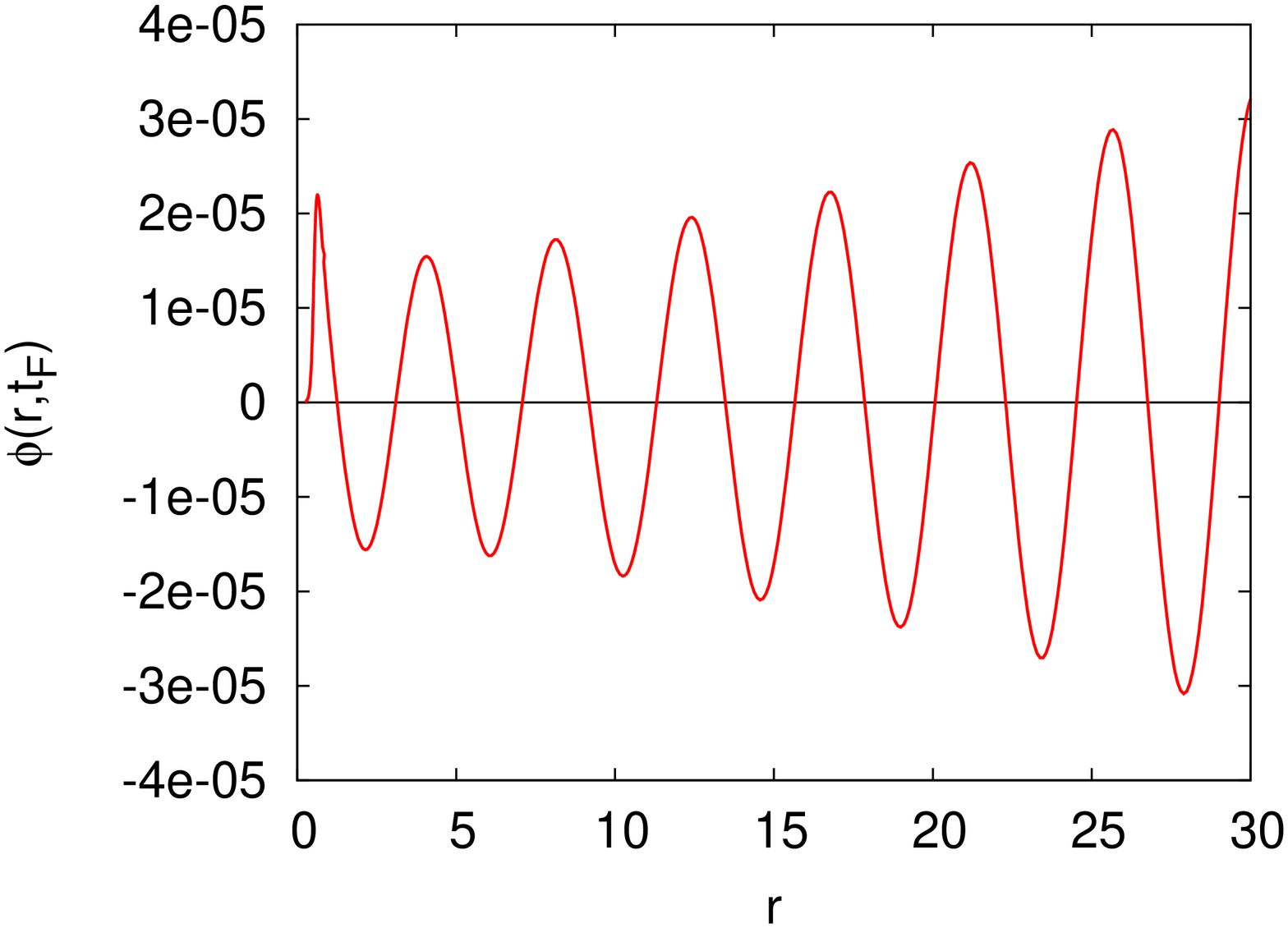}
  \includegraphics[angle=270,width=\linewidth]{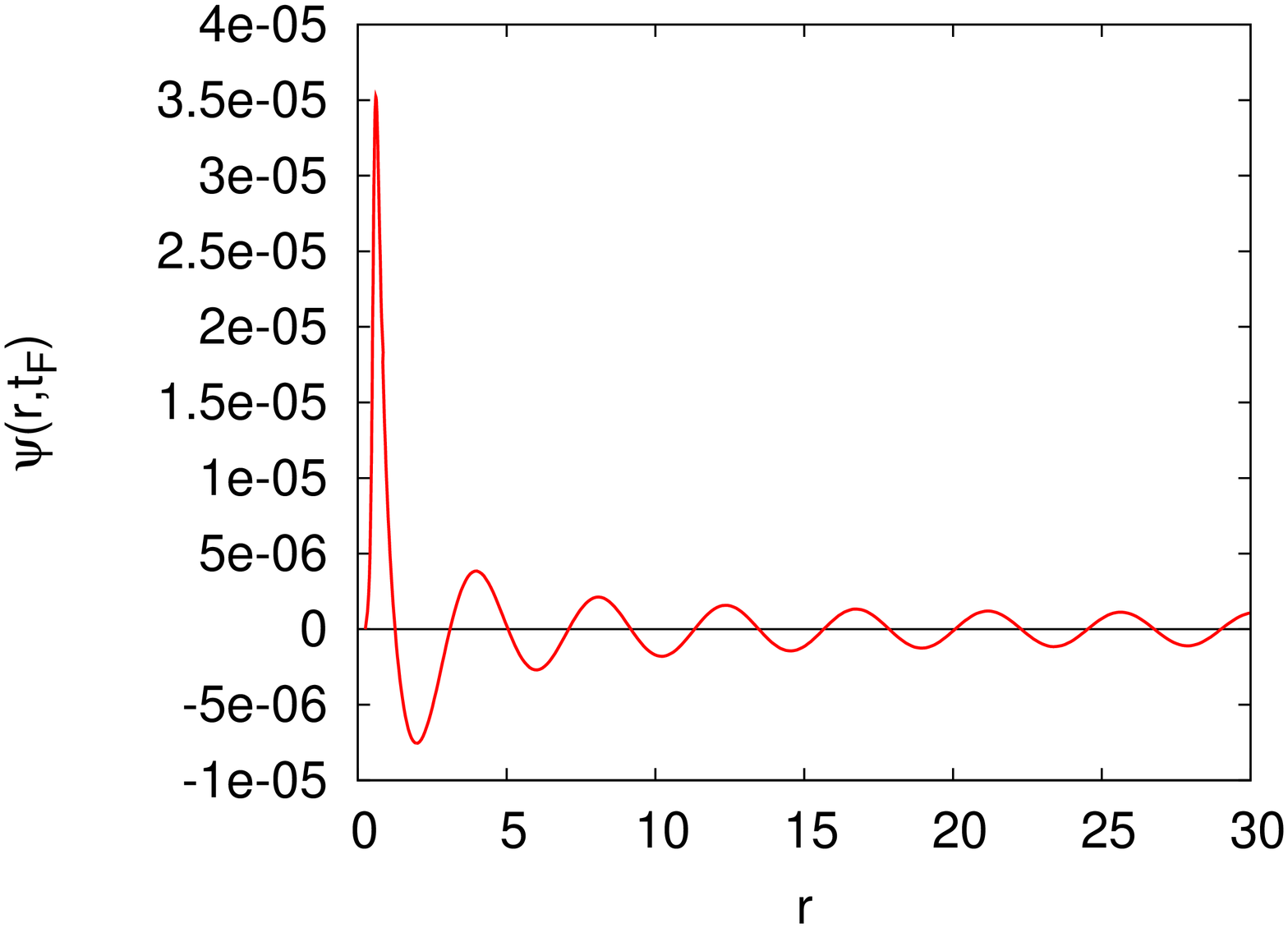}
\end{center}
\caption{Above: Behavior of $\phi$ with $l = 2$ as a function of $r$ near the center $r=0$ for a late time $t_F = 350$, shown here in order to exemplify the effect of conditions (\ref{BC1}) and (\ref{BC2}) in the numerical integration, for a spacetime with $q = 0.5$ and $q = 0.52$. Below: The same as in the left plot, but this time for the function $\psi = \phi/r$.}
\label{fig:phi_r}
\end{figure}

\begin{figure}[!htb]
\begin{center}
  \includegraphics[angle=270,width=\linewidth]{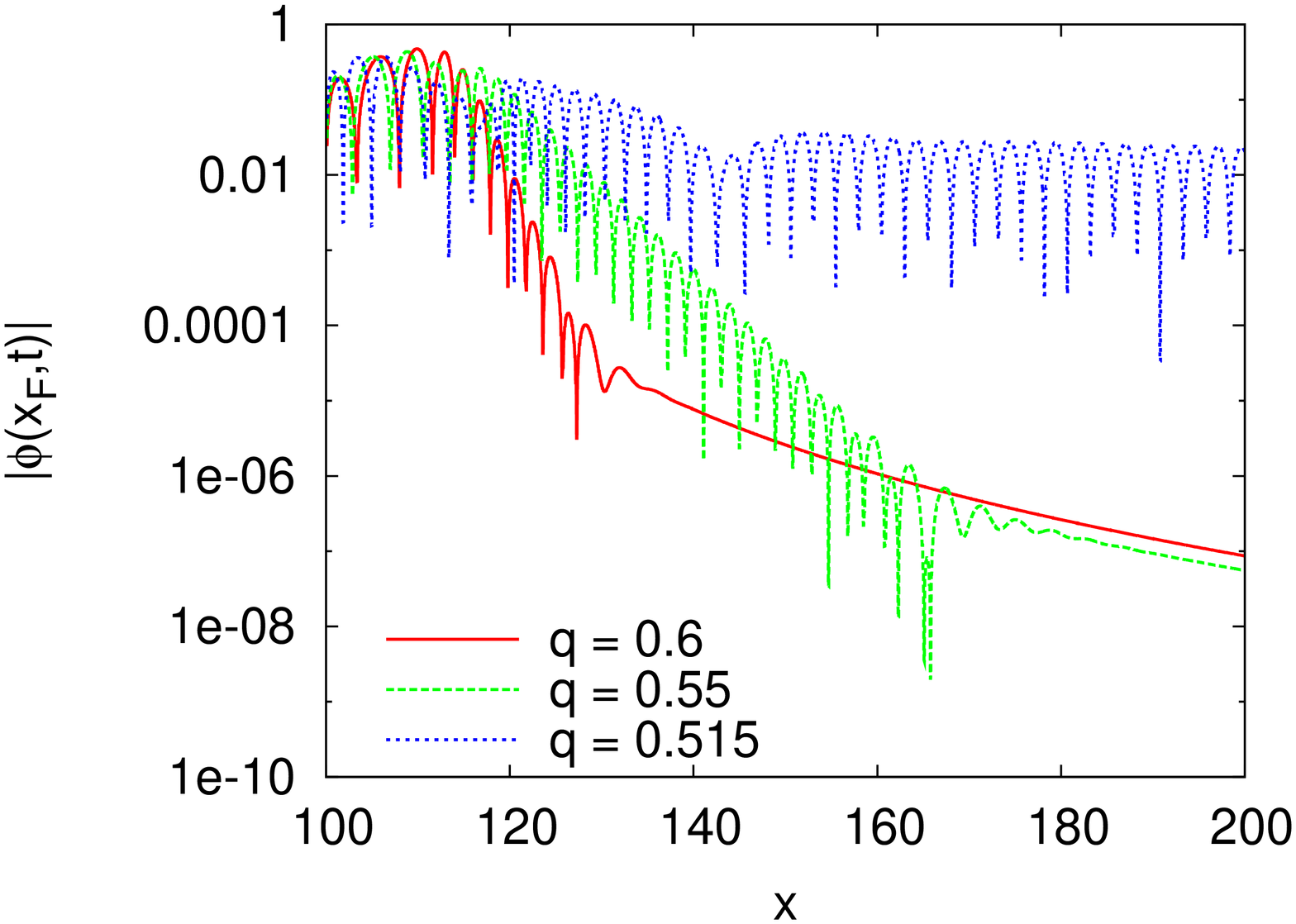}
  \includegraphics[angle=270,width=\linewidth]{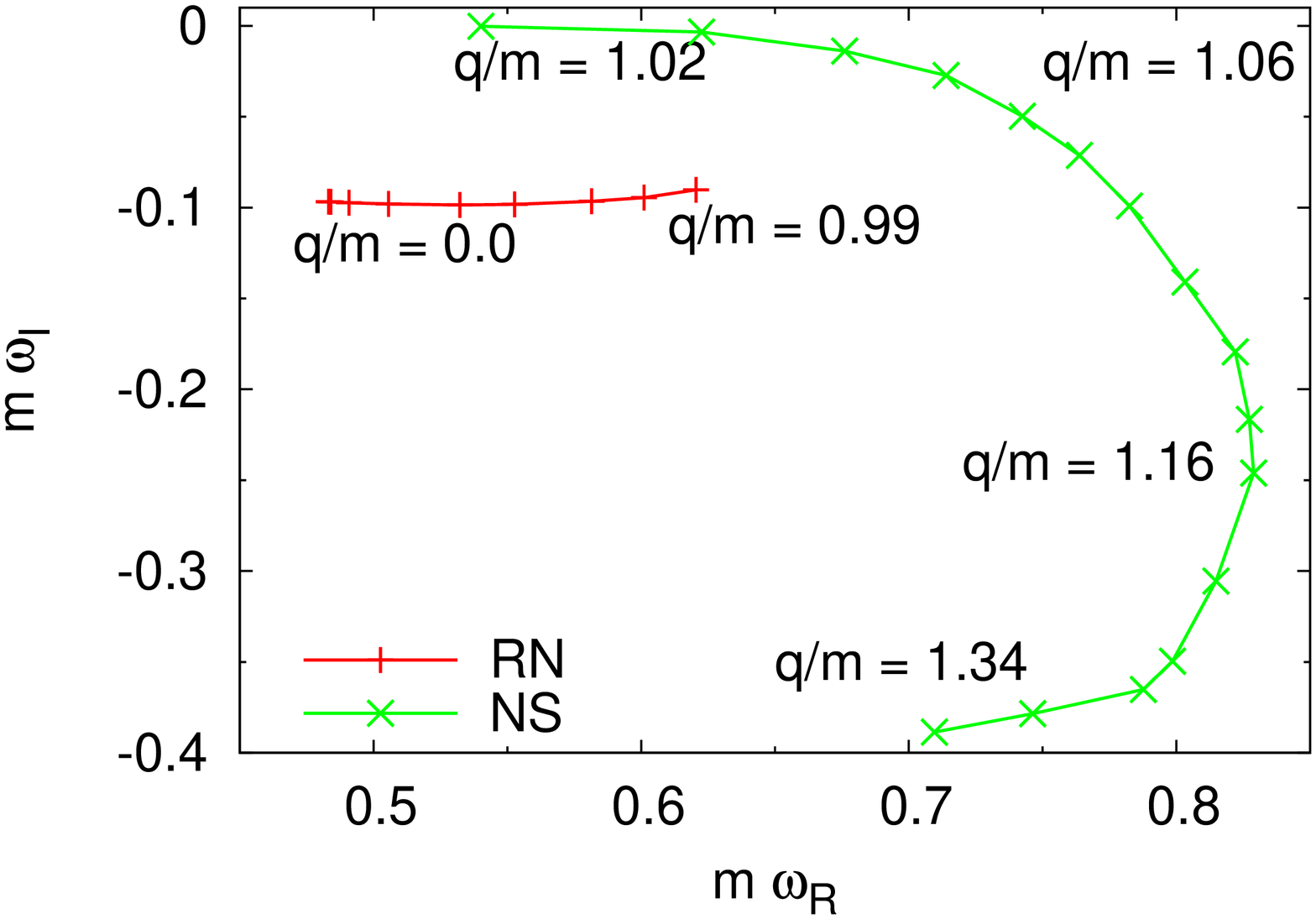}
\end{center}
\caption{Above: $\phi(x_F,t)$ with $l = 2$ at $x_F = 100$ for a spacetime with $m=0.5$ and different values of $q > m$. Below: Frequencies of the fundamental mode with $l = 2$ in the $\omega_R \times \omega_I$ plane, parametrized by the $q/m$ ration.}
\label{fig:field_qm}
\end{figure}

In the left plot of fig. \ref{fig:field_qm} we present some typical
time evolutions of $\phi$, for a $l = 2$ and different $q/m$ ratios.
In the right plot we present the obtained frequencies of the QNMs
(fundamental mode) in the $\omega_R \times \omega_I$ plane. We can
see a discontinuity in the frequencies as $q/m \to 1$, as was
expected from the discussion of the potential $V(r)$ (see fig.
\ref{fig:pot}). We also point here that we see no significant
changes, but rather a smooth behavior as $q^2/m^2 \to 9/8$ ($q/m \to
1.06$ in the plot). But we see a point of inflection in $\omega_R$
at $q/m \approx 1.16$, for which we did not find an analytical
explanation.

We remark here some numerical issues that prevented us from
extending the results shown in fig. \ref{fig:field_qm} for lower and
higher $q/m$ rations. As $q/m \to 1$ and $\phi$ becomes less damped,
different modes subsist for longer times and and longer evolutions
are needed in order to obtain the clear frequency of the fundamental
mode. On the other hand, as we increase $q/m$, $\phi$ is damped so
quickly that we cannot observe enough oscillation cycles to obtain
the frequencies. These are the issues that have limited our results.

\begin{figure}[!htb]
\begin{center}
  \includegraphics[angle=270,width=\linewidth]{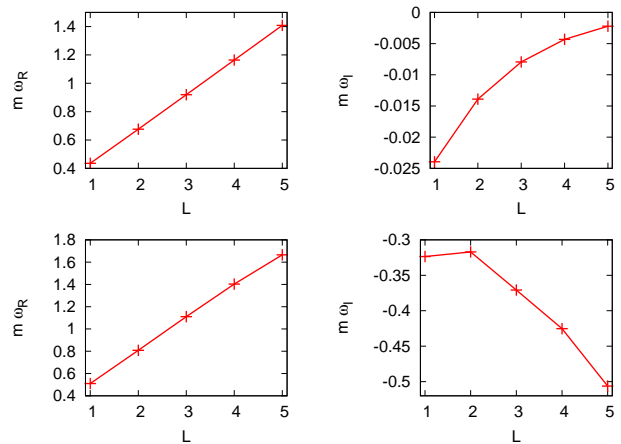}
\end{center}
\caption{Above: frequencies of the fundamental mode as a function of $l$ for $m = 0.5$ and $q = 0.52$ ($q^2/m^2 < 9/8$). Below: same as above, but this time for $m = 0.5$ and $q = 0.6$ ($q^2/m^2 > 9/8$). }
\label{fig:field_L}
\end{figure}

Finally, in fig. \ref{fig:field_L} we explore how the frequencies of
the QNMs change with $l$. As usual in black hole scattering
problems, we see that the oscillation frequency $\omega_R$ increases
with $l$. But the qualitative behaviour of $\omega_I$ changes
significantly with $q/m$. In the upper plots ($q^2/m^2 \lesssim
9/8$), $|\omega_I|$ decreases with $l$, that is, the damping time is
longer. In the lower plots ($q^2/m^2 \gtrsim 9/8$), we have the
opposite tendency. This behaviour is connected to the potential
$V(r)$ shown in fig. \ref{fig:pot_L}. It might be also interesting
to mention that in case $q^2/m^2 \gtrsim 9/8$, there is a
qualitative similarity in the behaviour of the imaginary part of the
frequencies as a function of $l$, between the case when $l$ is small
and the $l$ large limit.

\section{Asymptotically highly damped modes for the scalar field in the R-N naked singularity spacetime}

Let us explore the following limit, $l$ will be fixed and $n \to \infty$, assuming that there exist an infinite number of QNMs. Assume further that $\omega^{2}, ~\omega_{I}$ diverge as $n\to\infty$. 
In such a limit, to have a good understanding of what the QNM boundary conditions are, it is convenient to take an analytic continuation of the solution into the complex $x$ plane and impose the outgoing wave boundary conditions on the (Stokes) line $Im(\omega x)=0$. This is for almost imaginary asymptotic frequencies ($\omega\approx i \omega_{I}\to -i \infty$ as $n\to\infty$) the line  $Re(x)=0$.
The key point is to realize that, in the asymptotic case, $\omega^{2}\gg |V(x)|$ holds everywhere apart of a tiny region around $0$, where the potential $V(x)$ can be already approximated by 
\begin{equation}
V(x)\approx -\frac{2}{9x^{2}}~.
\end{equation}
(There is one subtlety: The potential, and also later the solutions, are multi-valued functions in the complex variable, but after specifying a branch cut together with the branch, this does not cause any problems to our analysis.) So let us make the following statement: The solutions of the equation 
\begin{equation}\label{high}
 \frac{d^{2}\phi_{l}(x)}{dx^{2}}+\left(\omega^{2}+\frac{2}{9 x^{2}}\right)\phi_{l}(x)=0
\end{equation}
are \emph{everywhere} good approximations to the solutions of the scalar field equation for the asymptotically highly damped modes. (The region where the  approximations $x\approx r^{3}$ and $V(x)\sim 1/r(x)^{6}$ cease to hold is the region where we can already neglect the potential as a whole with respect to the $\omega^{2}$ term.)

The equation \eqref{high} has a general solution given as 
\begin{equation}
\phi_{l}(x) = 
 \sqrt{\omega x} \left[A J_{1/6}(\omega x)+B J_{-1/6}(\omega x)\right]~,
\end{equation}
where $J_{a}$ is the Bessel function. The solution for which $\psi_{l}(x)$ fulfils the vanishing boundary condition at 0 is given by $B=0$.
Our QNM boundary condition says that this solution taken along the Stokes line  $Im(\omega x)=0$, should give purely outgoing radiation for $\omega x\to \infty$. But in the $\omega x\to\infty$ limit we know that we can approximate the solution (via Bessel functions approximations in such a limit) as
\begin{eqnarray} 
A \sqrt{\omega x}  J_{1/6}(\omega x)&\approx& A \sqrt{\frac{2}{\pi}} \cos\left(\omega x-\frac{\pi}{3}\right)=\\
&=&A \sqrt{\frac{1}{2\pi}}  \left(e^{-i\pi/3}  e^{i\omega x}+e^{i\pi/3}e^{-i\omega x}\right).\nonumber
\end{eqnarray}
This linear combination is independent on $\omega$ and does not give the purely outgoing radiation ($\sim e^{i\omega x}$), which suggests that our QNM boundary conditions cannot be fulfilled for asymptotically highly damped modes. 

This can be taken as an evidence that the asymptotically highly damped modes ($\omega_{I}\to -\infty$) do \emph{not} exist for the R-N naked singularity and this can be taken as a confirmation that the highly damped QNMs could link to the black hole spacetime horizon's properties (non-existence of the horizon leads to non-existence of the modes in the asymptotic limit). One can see also some link to the fact that for the extremal R-N black hole the spacing between the asymptotic frequencies goes to zero as proportional to the surface gravity (this might make a whole infinite tower of arbitrarily highly damped modes eventually ``collapse'' to a single mode). Moreover, we assume that the non-existence of the highly damped modes can be proven along the same lines also for another naked singularity spacetimes (like negative mass Schwarzschild spacetime etc \cite{Ishibashi}).

\section{Conclusions}

In this paper we analysed the problem of the scalar field scattering on
a R-N naked singularity background from the point of view of
quasi-normal modes. The evolution on the R-N naked singularity is
non-unique unless one specifies an additional boundary condition
representing a ``hair'' of the singularity. The quasi-normal modes
then carry information about the ``hair''. We applied a particular
boundary condition, that nothing comes out, or in from the
singularity and analysed analytically, as well as numerically, the
characteristic oscillations of the scalar field perturbations (low
damped quasi-normal modes). We analysed the eikonal $l\gg 1$ case via
the analytical approach confirming the intuition obtained through
the massless particle viewpoint, and showed that an approach based
on analytical approximations can be useful also for the small $l$
wave mode numbers. For the small $l$-s we calculated the frequencies
numerically via the characteristic integration method. We also suggested arguments showing that the asymptotically highly damped modes (limit $l$ fixed and $n\to \infty$) do \emph{not} exist in case of R-N naked singularity. This might confirm the intuition one has about such modes from the black hole physics (and from the presently popular conjectures
\cite{Hod,Maggiore}).

The basic results can be summarized as follows: for the low modes and the large $l$ there is a continuous transition in the low damped QNM modes between the R-N black hole and the R-N naked singularity. However, when the ratio $q^{2}/m^{2}$ becomes larger than approximately $9/8$ then the picture becomes significantly different and the low damped modes do \emph{not} exist for large $l$-s. (This is a very different picture from the BH based intuition.) For the small $l$ numbers the modes face a discontinuous transition when transiting from the black hole to the naked singularity. Furthermore, the $l$ dependence $|\omega_I|$ (for small $l$) changes as $q^{2}/m^{2}$ becomes larger than approximately $9/8$: $|\omega_I|$ decreases for $q^{2}/m^{2} \lesssim 9/8$ and increases for $q^{2}/m^{2} \gtrsim 9/8$. It might be interesting to notice that for $q^{2}/m^{2} \gtrsim 9/8$ the increase of $|\omega_{I}|$ as a function of $l$ (for small $l$-s) matches the behaviour of $|\omega_{I}|$ for large $l$-s. In the case of large $l$-s and $q^{2}/m^{2} \gtrsim 9/8$ we have shown that $|\omega_{I}|$ of the fundamental mode grows at least cubically with $l$ and thus, as we already mentioned, the low damped modes do not exist. For the asymptotically highly damped modes our results seem to suggest that they do not exist, which means that the imaginary parts of the frequencies are bounded.

\subsection*{Acknowledgments} The authors are grateful to FAPESP, CNPq, and the Max Planck Society for the financial
support.

\appendix

\section{Calculations of some of the important quantities\label{B}}
In this appendix section we provide a list of some of the quantities (also some of their derivations) that occur in the calculations relevant to the problem analysed in this paper.
Write $V(r)$ in a convenient form:
\begin{equation}
V(r)=\frac{A}{r^{2}}+\frac{B}{r^{3}}+\frac{C}{r^{4}}+\frac{D}{r^{5}}+\frac{E}{r^{6}},
\end{equation}
with
\begin{eqnarray}
A&\equiv &l(l+1),\\
B&\equiv &2m[1-l(l+1)],\\
C&\equiv &q^{2}l(l+1)-(2m)^{2}-2q^{2},\\
D&\equiv &6mq^{2},\\
E&\equiv &-2q^{4}.
\end{eqnarray}
Now let us calculate the second derivative of the potential with
respect to the tortoise coordinate:
\begin{equation}
\frac{d^{2}V(x)}{dx^{2}}=\frac{d^{2}V(r)}{dr^{2}}\left(\frac{dr}{dx}\right)^{2}+\frac{dV(r)}{dr}\frac{d^{2}r}{dx^{2}}.
\end{equation}
At $r_{max}$ the derivative $dV(r)/dr$ is zero, then necessarily
\begin{equation}
\frac{d^{2}V(x)}{dx^{2}}_{|x_{max}}=\left[\frac{d^{2}V(r)}{dr^{2}}\left(\frac{dr}{dx}\right)^{2}\right]_{|x_{max}}
\end{equation}
and since
\begin{equation}
\frac{d^{2}V(r)}{dr^{2}}=\frac{6A}{r^{4}}+\frac{12B}{r^{5}}+\frac{20C}{r^{6}}+\frac{30D}{r^{7}}+\frac{42E}{r^{8}}
\end{equation}
we obtain
\begin{eqnarray}
\frac{d^{2}V(r)}{dx^{2}}&=&\left(1-\frac{2m}{r}+\frac{q^{2}}{r^{2}}\right)^{2}\\
&\times &
\left(\frac{6A}{r^{4}}+\frac{12B}{r^{5}}+\frac{20C}{r^{6}}+\frac{30D}{r^{7}}+\frac{42E}{r^{8}}\right).\nonumber 
\end{eqnarray}
 For the global extremum condition holds:
\begin{equation}\label{extremum}
\frac{dV(r)}{dr}=2Ar_{max}^{4}+3Br_{max}^{3}+4Cr_{max}^{2}+5Dr_{max}+6E=0.
\end{equation}

This equation can have maximally 4 roots. From \eqref{potential} it
can be easily seen that for $r\to\infty$ \eqref{potential} becomes
for any $l,q,m$ positive (dominant term is $l(l+1)/r^{2}$) and goes
to 0. In case of naked singularity for $r\to 0$ the potential goes
always to $-\infty$ (dominant term is $-2q^{4}/r^{6}$), in case of
black hole for $r\to r_{+}$ it always goes to 0 (dominant term is
$f(r)$). This means, together with \eqref{extremum}, that for any
$m,q,l$ (black hole or not) there are always either two local maxima
and one local minimum, or one local maximum without local minima.

\subsection{The large $l$ limit}
For $l\gg 1$
the following approximations hold: For large $l$ the maximum of the
inner peak can be approximately found analytically and behaves as:
\begin{equation}
r_{max}\approx\frac{\sqrt{3}|q|}{\sqrt{l(l+1)}}.
\end{equation}
Also holds the following: For $x\ll 1$
\begin{equation}\label{relationtortoise}
x\approx \frac{r^{3}}{3q^{2}}.
\end{equation}
The parameters $\alpha_{1}, V_{1}$ turn in the large $l$ limit to
be:
\begin{equation}
\alpha_{1}\approx \frac{2}{3}\frac{[l(l+1)]^{3/2}}{|q|},
\end{equation}
\begin{equation}
\sqrt{V_{1}}\approx \frac{[l(l+1)]^{3/2}}{3^{3/2} |q|}.
\end{equation}
This implies also the following results: For $l\to\infty$
\begin{eqnarray}
x_{1max}\alpha_{1} &\to & \frac{2}{\sqrt{3}},\\
\frac{\sqrt{V_{1}}}{\alpha_{1}} &\approx & \frac{1}{2\sqrt{3}},\\
V_{2}& \thicksim & l(l+1),\\
\frac{V_{2}}{V_{1}} &\to & 0,\\
\frac{\sqrt{V_{2}}}{\alpha_{1}} &\to & 0,\\
\frac{\sqrt{V_{2}}}{\alpha_{2}}&\to& \infty,\\
V(x) &\approx& -\frac{2}{9x^{2}} ~~\hbox{for}~~~~x\ll 1.
\end{eqnarray}

\section{Some suggestions for the analytical treatment of the scattering for small wave mode numbers\label{A}}

Write the two linearly independent solutions on the domain $i$ in
the form convenient for the logarithmic derivative gluing condition:
$C_{i}(\Psi_{i1}+K_{i}\Psi_{i2})$. If the domain extends to the
infinity and $\Psi_{i1}$ is taken to be the asymptotically incoming
wave solution and $\Psi_{i2}$ the asymptotically outgoing wave
solution, then $K_{i}(\omega)$ is the S-matrix (ratio of the
coefficients of the outgoing and incoming waves). Anyway, the
logarithmic derivative gluing condition at the boundary of the
regions $i$ and $i-1$ turns to be:
\begin{widetext}
\begin{equation}\label{gluing}
K_{i}=\left[\frac{\Psi'_{(i-1)1}\Psi_{i 1}-\Psi'_{i
1}\Psi_{(i-1)1}+K_{i-1}\left(\Psi'_{(i-1)2}\Psi_{i 1}-\Psi'_{i
1}\Psi_{(i-1)2}\right)}{\Psi'_{i 2}\Psi_{(i-1) 1}-\Psi'_{(i-1)
1}\Psi_{i 2}+K_{i-1}\left(\Psi'_{i 2}\Psi_{(i-1) 2}-\Psi'_{(i-1)
2}\Psi_{i 2}\right)}\right]_{|_{a_{i}}}
\end{equation}
\end{widetext}
This is the condition connecting the $K_{i}$ coefficient with
$K_{i-1}$ coefficient through the values of the solutions and their
derivatives at the boundary of the regions. On one side consider asymptotically free-wave region, whereas on the other side consider infinite potential barrier-like boundary condition for $\Psi$. In such case one can
through equations of the type \eqref{gluing} connect (after finite
number of steps) the S-matrix of the external (asymptotic) region
with the boundary condition at the origin (where
we can conveniently put $K_{1}$ to be zero). Then one gets the
S-matrix expressed through a function of the boundary values (at all
the boundaries) of the solutions and their derivatives. This will be
algebraically more difficult condition than the one coming just from
posing the outgoing wave condition in the external region, as we did
in the previous sections (case of large $l$ and $q^{2}/m^{2}$ larger
than approximately 9/8). The reason for posing the more complicated
condition of the type \eqref{gluing} is that it might be easier (and
also brings more insight) to look for a real energy resonances (by
plotting phase change and looking for rapid phase shifts, hence
rapid change of the logarithm of the S-matrix), than to try to
numerically solve the more simple outgoing wave condition (with the help of appropriate computer software). Such resonances should then 1 to 1 correspond to the
low damped quasi-normal frequencies. Let us apply these ideas and try to calculate the expressions for the partial S-matrix.

\subsection{The case for $q^{2}/m^{2}\gtrsim 9/8$}
This is the case \ref{one}. We are gluing only two
regions. The solution on the left side of the boundary $\Psi_{L}$,
such that fulfils $\Psi_{L}(0)=0$, is again given as \eqref{modes2}.
On the right side (the Morse potential side) the solution is
given as:
\begin{widetext}
\begin{equation}\label{Morse}
\Psi_{R}(x)=C_{1}[\Psi_{R/I}(x)+S(\omega,\alpha_{1},V_{1})\Psi_{R/O}(x)]= 
C_{1}e^{-\frac{z}{2}}\left[
e^{-i\omega x} M(s_{1},s_{2},z)+S(\omega,\alpha_{1},V_{1})
e^{i\omega x}M(s'_{1},s'_{2},z)\right].
\end{equation}
By $\Psi_{R/O}(x),\Psi_{R/I}(x)$ we mean asymptotically outgoing and
incoming wave solutions in the sense that $\Psi_{R/I}(x\to\infty)\to
e^{-i\omega x}$ and $\Psi_{R/O}(x\to\infty)\to e^{i\omega x}$
exactly. Furthermore $M(.,.,.)$ is being the Kummer (confluent)
hypergeometric function and
\begin{eqnarray}
s_{1}&\equiv&\frac{1}{2}-i\frac{\sqrt{V_{1}}}{\alpha_{1}}+\frac{i\omega}{\alpha_{1}}\equiv s'_{1}+\frac{2i\omega}{\alpha_{1}},\\
s_{2}&\equiv& 1+\frac{2i\omega}{\alpha_{1}}\equiv s'_{2}+\frac{4i\omega}{\alpha_{1}},\\
z&\equiv& \frac{i2\sqrt{V_{1}}}{\alpha_{1}}\exp(-\alpha_{1}(x-x_{1max})).
\end{eqnarray}
Also $S_{l}(\omega,\alpha_{1},V_{1})$ is the S-matrix. The
logarithmic derivatives gluing condition at
$a_{1}=x_{1max}-\frac{1}{\alpha_{1}}\ln(2)$ is then
\begin{equation}\label{onepeak}
f(a_{1})\left[\frac{1}{a_{1}}+\frac{\beta_{1}e^{\beta_{1}a_{1}}-\beta_{2}e^{\beta_{2}a_{1}}}{e^{\beta_{1}a_{1}}-e^{\beta_{2}a_{1}}}\right]= 
\frac{\alpha_{1}z(a_{1})}{2}+i\omega+\frac{s'_{1}}{s'_{2}}\frac{M\left(s'_{1}+1,
s'_{2}+1,z(a_{1})\right)}{M\left(s'_{1}, s'_{2},z(a_{1})\right)}.
\end{equation}
Considering that the following holds:
\begin{equation}z(a_{1})=\frac{i4\sqrt{V_{1}}}{\alpha_{1}},\end{equation}
one can rewrite \eqref{onepeak} into:
\begin{equation}\label{onepeak2}
f(a_{1})\left[\frac{1}{a_{1}}+\frac{\beta_{1}e^{\beta_{1}a_{1}}-\beta_{2}e^{\beta_{2}a_{1}}}{e^{\beta_{1}a_{1}}-e^{\beta_{2}a_{1}}}\right]= 
i2\sqrt{V_{1}}+i\omega+\frac{s'_{1}}{s'_{2}}\frac{M\left(s'_{1}+1,
s'_{2}+1,\frac{i4\sqrt{V_{1}}}{\alpha_{1}}\right)}{M\left(s'_{1},
s'_{2},\frac{i4\sqrt{V_{1}}}{\alpha_{1}}\right)}.
\end{equation}
\end{widetext}
The partial S-matrix can be obtained from \eqref{gluing} ($K_{1}=0$)
as:
\begin{equation}\label{gen}
S_{l}(\omega, V_{1},
\alpha_{1})=-\frac{\Psi_{R/I}}{\Psi_{R/O}}\frac{\frac{\Psi'_{L}}{\Psi_{L}}-\frac{\Psi'_{R/I}}{\Psi_{R/I}}}{\frac{\Psi'_{L}}{\Psi_{L}}-\frac{\Psi'_{R/O}}{\Psi_{R/O}}}.
\end{equation}
Since for real $\omega$ holds that $\Psi_{R/O}=\Psi_{R/I}^{*}$ and
$\Psi_{L}$ is a real valued function, one can immediately observe
that (for $\omega\in\mathbb{R}$) $S_{l}^{*}(\omega)=S^{-1}_{l}(\omega)$,
hence partial S-matrix is unitary.

From \eqref{gen} we can conclude the following
\begin{widetext}
\begin{eqnarray}\label{Smatrix}
S_{l}(\omega, V_{1}, \alpha_{1})&=&-e^{-2i\omega
a_{1}}\frac{M\left(s_{1},s_{2},\frac{i4\sqrt{V_{1}}}{\alpha_{1}}\right)}{M\left(s'_{1},s'_{2},\frac{i4\sqrt{V_{1}}}{\alpha_{1}}\right)} \\
&\times&
\frac{f(a_{1})\left[\frac{1}{a_{1}}+\frac{\beta_{1}e^{\beta_{1}a_{1}}-\beta_{2}e^{\beta_{2}a_{1}}}{e^{\beta_{1}a_{1}}-e^{\beta_{2}a_{1}}}\right]-\left[i2\sqrt{V_{1}}-i\omega+\frac{s_{1}}{s_{2}}\frac{M\left(s_{1}+1,
s_{2}+1,\frac{i4\sqrt{V_{1}}}{\alpha_{1}}\right)}{M\left(s_{1},
s_{2},\frac{i4\sqrt{V_{1}}}{\alpha_{1}}\right)}\right]}{f(a_{1})\left[\frac{1}{a_{1}}+\frac{\beta_{1}e^{\beta_{1}a_{1}}-\beta_{2}e^{\beta_{2}a_{1}}}{e^{\beta_{1}a_{1}}-e^{\beta_{2}a_{1}}}\right]-\left[i2\sqrt{V_{1}}+i\omega+\frac{s'_{1}}{s'_{2}}\frac{M\left(s'_{1}+1,
s'_{2}+1,\frac{i4\sqrt{V_{1}}}{\alpha_{1}}\right)}{M\left(s'_{1},
s'_{2},\frac{i4\sqrt{V_{1}}}{\alpha_{1}}\right)}\right]} \nonumber
\end{eqnarray}

\subsection{The two peak case, $q^{2}/m^{2} \lesssim 9/8$\label{section}}
The two peak case is a more complicated problem. Anyway, one can write the logarithmic derivative gluing
condition as
\begin{eqnarray}\label{equation1}
\frac{\alpha_{1}z(a_{2})}{2}&-&i\omega\frac{S_{1}(\omega,\alpha_{1},V_{0}) M\{s_{1},s_{2},z(a_{2})\}+e^{i2\omega a_{2}}M\{s'_{1},s'_{2},z(a_{2})\}}{S_{1}(\omega,\alpha_{1},V_{0}) M\{s_{1},s_{2},z(a_{2})\}-e^{i2\omega a_{2}}M\{s'_{1},s'_{2},z(a_{2})\}}\\ 
&+&
\frac{S_{1}(\omega,\alpha_{1},V_{0})
M\{s_{1}+1,s_{2}+1,z(a_{2})\}\frac{s_{1}}{s_{2}}- e^{i2\omega
a_{2}}M\{s'_{1}+1,s'_{2}+1,z(a_{2})\}\frac{s'_{1}}{s'_{2}}}{S_{1}(\omega,\alpha_{1},V_{0})
M\{s_{1},s_{2},z(a_{2})\}- e^{i2\omega
a_{2}}M\{s'_{1},s'_{2},z(a_{2})\}} \nonumber\\
&=&i\omega\frac{F_{21}\{g_{1},g_{2},g_{3}-1,
[1+\exp(2\alpha_{2}(a_{2}-x_{2max}))]^{-1}\}}{F_{21}\{g_{1},g_{2},g_{3},
[1+\exp(2\alpha_{2}(a_{2}-x_{2max}))]^{-1}\}}. \nonumber
\end{eqnarray}
Here $S_{1}$ is the S-matrix from the case \ref{one} and
is given by the formula \eqref{Smatrix}. The S-matrix related to the
case \ref{three} (call it $S_{2}$) is related to $S_{1}$ through the
condition \eqref{gluing}. In the gluing formula \eqref{gluing}
$S_{2}$ stands for $K_{i}$ and $S_{1}$ for $K_{i-1}$. Also $\Psi_{i
1,2}$ are the solutions of the Poeschl-Teller case
\begin{equation}
\Psi_{i1}(x)=e^{i\omega x}F_{21}\left[g_{1},g_{2},g_{3},
[1+\exp(2\alpha_{2}(x-x_{2max}))]^{-1}\right],
\end{equation}
and
\begin{equation}
\Psi_{i2}(x)=e^{-i\omega x}F_{21}\left[g_{1},g_{2},g_{3},
[1+\exp(-2\alpha_{2}(x-x_{2max}))]^{-1}\right].
\end{equation}
\end{widetext}
$\Psi_{(i-1)1,2}$ are the Morse potential solutions given as
\eqref{Morse}. We see that if there is a regime in which the modes
do not feel the smaller peak, then they are given by the poles of
$S_{1}$ and must be in the same time poles of $S_{2}$. The formula
for $S_{2}$ in this case (for frequencies that represent poles of
$S_{1}$) simplifies to
\begin{equation}\label{Smatrix2}
S_{2}=\left[\frac{\left(\Psi'_{(i-1)2}\Psi_{i 1}-\Psi'_{i
1}\Psi_{(i-1)2}\right)}{\left(\Psi'_{i 2}\Psi_{(i-1) 2}-\Psi'_{(i-1)
2}\Psi_{i 2}\right)}\right]_{|_{a_{i}}}.
\end{equation}
There is a possibility that for the case of the low modes in the two
peak model one might be able to neglect the infinite depth of the
valley between the infinite barrier and the Morse peak. In such case
one can take a simpler model where the Morse potential directly
follows after the infinite barrier. The $S_{1}$ term appearing in
the formula \eqref{equation1} will be in such case given by a much
more simple expression:
\begin{equation}
S_{1}(\omega,\alpha_{1},V_{0})=\frac{M\{s_{1},s_{2},z(0)\}}{M\{s'_{1},s'_{2},z(0)\}}.
\end{equation}

\end{document}